\documentclass[twocolumn,nonbibnotes,nofootinbib,notitlepage,longbibliography]{revtex4-2}
\usepackage[utf8]{inputenc}
\usepackage[T1]{fontenc}
\usepackage{amsmath}
\usepackage{bm}
\usepackage{graphicx}
\usepackage{bm}
\usepackage{epsfig}
\usepackage{epstopdf}
\usepackage{amsmath}
\newcommand{\divv}{\mathop\mathrm{div}}
\usepackage{color}
\usepackage{xcolor}
\usepackage[colorlinks=true,citecolor=green!60!black,linkcolor=blue]{hyperref}
\usepackage{orcidlink}

\begin{document}
	
\title{Long-range exchange interaction controls the fine structure \\of excited trion states in semiconductor quantum dots}
	\author{M.M. Glazov\orcidlink{0000-0003-4462-0749}}\email{glazov@coherent.ioffe.ru}
	\author{E.L. Ivchenko\orcidlink{0000-0001-7414-462X}}%\email{ivchenko@coherent.ioffe.ru}
	\affiliation{Ioffe Institute, 194021 St. Petersburg, Russia}

	\begin{abstract}
We develop a microscopic theory of the long-range electron-hole exchange interaction in charged excitons (trions)
confined in semiconductor quantum dots. While the ground-state singlet trion remains degenerate in the spin 
component of unpaired charge carrier by time-reversal symmetry, excited trion states exhibit a rich fine structure 
resulting from the interplay of electron-electron and electron-hole exchange interactions. We derive the effective 
long-range exchange Hamiltonian for both the spin-$3/2$ heavy-hole and a simple spin-$1/2$ valence band models. The 
long-range exchange interaction mixes singlet and triplet trion configurations, giving rise to anisotropic fine-structure splittings and corresponding linear-polarization-dependent optical spectra determined by the quantum-dot shape. Conversely, the short-range exchange interaction is isotropic and does not lead to linear polarization of trion-related transitions.
Analytical expressions are obtained for the long-range exchange parameters. The developed theory establishes a 
unified microscopic description of the fine structure of excited trions in semiconductor quantum dots and provides a
framework for interpreting polarization-resolved optical spectroscopy of charged excitonic complexes. 
		\end{abstract}
	%\pacs{71.35.--y, 71.36.+c, 78.30.--j, 81.05.Hd}
	
	\date{\today}
	\maketitle

\section{Introduction}\label{sec:intro}

The fine structure of excitonic states underlies a wide range of optical and spin-dependent phenomena, including reflection, absorption, emission, and scattering of light, coherent spin dynamics, and the generation of entangled photon pairs. The theory of fine structure of Wannier--Mott excitons has a long history. The development of the theory proceeded from bulk materials \cite{dresselhausEffectiveMassApproximation1956,rashbaEffectResonanceExcitation1959,pikusExchangeInteractionExcitons1971,Kiselev,
denisovLongitudinalTransverseExcitons1973,birSymmetryStraininducedEffects1974} to two-dimensional excitons \cite{andreaniRADIATIVELIFETIMEFREE1991,Andreani2,maialleExcitonSpinDynamics1993,glazovExcitonFineStructure2014} to zero-dimensional excitons localized in quantum well structures \cite{gupalovFineStructureLocalized1998} or confined in quantum dots \cite{ivchenko151,franceschettiShortrangeLongrangeElectronhole1998,ivchenko231,ivchenko232,Hawrylak}. By now, the microscopic theory of short-range and long-range exchange (LRE) interaction of electrons and holes in neutral excitons is well established and provides a quantitative description of fine-structure splittings, optical selection rules, and exciton spin dynamics in semiconductor nanostructures. Detailed understanding of the exchange interaction enabled efficient control of the exciton fine structure \cite{stevensonMagneticfieldinducedReductionExciton2006,kowalikInfluenceInplaneElectric2005,
ivchenko233} and paved the way to an efficient realization of photon entanglement in semiconductor nanosystems~\cite{akopianEntangledPhotonPairs2006,stevensonSemiconductorSourceTriggered2006,
dousseUltrabrightSourceEntangled2010,rakhlinDemultiplexedSinglephotonSource2023,
wuPurcellenhancedTwophotonEmission2026}.

M.A. Lampert \cite{lampertMobileImmobileEffectiveMassParticle1958} extended the concept of the neutral exciton to consider charged excitonic complexes containing two or more identical charge carriers. For the charged three-particle excitonic complexes with two electrons and a hole or two holes and an electron, he coined the term ``excitonic ions'' $--+$ and $++-$. These complexes are now universally known as negatively ($X^-$) and positively ($X^+$) charged excitons, or trions. The first observation of charged excitons $X^{-}$ was achieved in 1993 by Kheng et al. in an $n$-doped II-VI quantum well structure \cite{khengObservationNegativelyCharged1993}. It was followed by observation of trions in III-V semiconductor quantum wells~\cite{finkelsteinNegativelyPositivelyCharged1996} and more recently in extremely thin materials~\cite{makTightlyBoundTrions2013,rossElectricalControlNeutral2013}. At present the ground state of trions has been extensively studied both in quantum wells and quantum dots, e.g. \cite{brackerBindingEnergiesPositive2005,liSubmillielectronvoltLineWidths2024}. Spatially resolved photoluminescence excitation spectroscopy has been performed to study the excited states of trions in single quantum dots \cite{jovanovDirectObservationMetastable2011,bennyTwophotonPhotoluminescenceExcitation2011,
bennyExcitationSpectroscopySingle2012,molasIntershellExchangeInteraction2013,
molasMagneticFieldEffect2014,smolenskiFineStructureResonantly2016,ivchenko302,
podemskiSingleDotPhotoluminescence2019,galimovChargeStatesSingle2025,
sorokinMetamorphicInAsInGaAs2025}. 
 
In rapidly developing field of quantum technologies the trions play special role in realization of spin-photon interfaces and entanglement \cite{schonSequentialGenerationEntangled2005,lindnerProposalPulsedOnDemand2009,
huDeterministicPhotonEntangler2008,arnoldMacroscopicRotationPhoton2015,
coganDeterministicGenerationIndistinguishable2023a,costeHighrateEntanglementSemiconductor2023,
suContinuousDeterministicAllphotonic2024,serovHiddenAnisotropyControls2025}. Thus, the fine structure of trion states is of high interest because it controls the spin-photon interface in charged quantum dots. While the ground-state trions do not show any fine structure splitting in the absence of magnetic field because of the time-reversal symmetry~
\cite{bayerFineStructureNeutral2002,iakovlevFermiPolaronFine2023}, the excited states of the trion can possess non-trivial fine structure as shown by K.V. Kavokin \cite{kavokinFineStructureQuantumdot2003} using the method of invariants. However, a microscopic theory explicitly relating the fine-structure parameters to the long-range electron-hole exchange interaction is absent to the best of our knowledge.

In the present work we develop a theory of LRE interaction and fine structure of excited trions in single quantum dots. A particular attention is focused on a trion complex formed by a resident 1$s$ electron  (occupying the ground state orbital) and 2$p$-2$p_h$ photoexcited electron-hole pair (occupying first excited states). The theory provides an effective LRE Hamiltonian expressed through the trion envelope functions and interband momentum matrix elements, predicts the fine-structure spectrum and polarization properties of the optical transitions. In Sect. \ref{sec:models} we introduce two models of semiconductor conduction and valence bands and present the singlet and triplet wavefunctions for the $X^{-*}$ trion of the $1s2p$-2$p_h$ configuration. The LRE interaction Hamiltonian is derived in Sect. \ref{sec:HLR}. The excited trions in the chosen two band-structure models are considered respectively in Sects.~\ref{subsec:hh:tr} and \ref{simple bands} followed by a brief discussion and concluding remarks in Sect.~\ref{sec:disc}.

\section{Trion states in quantum dots}\label{sec:models}

This section introduces the basics of the band structure and optical selection rules, Sec.~\ref{subsec:bands}, as well as involved single and few-particle states, Sec.~\ref{subsec:states}. These results are the key input for the analysis of the excited-trion fine structure related to the LRE interaction.

\subsection{Band structure models}\label{subsec:bands}
We apply two models of semiconductor quantum dots. The first one (model I) is the `simple' band model where both the conduction-band electrons and valence-band holes are characterized by the effective angular momentum 1/2. As an example, such simple band structure can be realized for the optical transitions involving the spin-orbit split-off $\Gamma_7$ valence band and the conduction $\Gamma_6$ band in zinc-blende-lattice-based semiconductor nanosystems \cite{birSymmetryStraininducedEffects1974,ivchenko135,ivchenko151,kudlacikLandegFactors2025}. Another example is the perovskite nanocrystals where the valence band is $\Gamma_6^+$ and the conduction band is $\Gamma_7^-$ \cite{beckerBrightTripletExcitons2018,ivchenko334}. For definiteness we consider the former case of the exciton $\Gamma_6 \times \Gamma_7$ and take the Bloch amplitudes in the form
\begin{subequations}
\label{Bloch:67}
\begin{align}
&|\Gamma_6, + 1/2 \rangle = \mathcal S \hspace{0.5 mm}| \hspace{-0.5 mm}\uparrow \hspace{0.5 mm} \rangle, \label{Bloch:67a}\\
&|\Gamma_6, -1/2\rangle = \mathcal S \hspace{0.5 mm} | \hspace{-0.5 mm} \downarrow \hspace{0.5 mm} \rangle, \label{Bloch:67b}\\
&|\Gamma_7, + 1/2\rangle = - \frac{1}{\sqrt{3}}\left[ \left( \mathcal X + \mathrm i \mathcal Y \right)  \hspace{0.5 mm} |  \hspace{-0.5 mm} \downarrow\rangle + \mathcal Z \hspace{0.5 mm} |  \hspace{-0.5 mm} \uparrow \hspace{0.5 mm} \rangle \right], \label{Bloch:67c}\\
&|\Gamma_7, - 1/2\rangle = -  \frac{1}{\sqrt{3}}\left[ \left( \mathcal X - \mathrm i \mathcal Y \right) \hspace{0.5 mm}| \hspace{-0.5 mm} \uparrow \hspace{0.5 mm} \rangle - \mathcal Z \hspace{0.5 mm}| \hspace{-0.5 mm}\downarrow \hspace{0.5 mm} \rangle  \right]. \label{Bloch:67d}
\end{align}
\end{subequations}
The amplitudes $\mathcal S \equiv \mathcal S(\bm r)$, $\mathcal X$, $\mathcal Y$, and $\mathcal Z$ transform as corresponding coordinates~\cite{birSymmetryStraininducedEffects1974,ivchenko135}, and $| \hspace{-0.5 mm} \uparrow \hspace{0.5 mm} \rangle, | \hspace{-0.5 mm} \downarrow \hspace{0.5 mm} \rangle$ are the spin-up and spin-down columns. With this definition of the Bloch amplitudes, the $\sigma^+$ circularly polarized light propagating along the $z$ axis induces the transition between the states $|\Gamma_7,-1/2\rangle$ in the valence band and $|\Gamma_6,+1/2\rangle$ in the conduction band, i.e., an electron and a hole with spin projections $+1/2$ on the $z$ axis are generated as a result of the $\sigma^+$ photon absorption. In the $\sigma^-$ polarization, the transition creates a pair with the spin projections of $-1/2$. The $z$-polarized light creates electron-hole pairs with the spin components $(s_e,s_h)$ being $(\pm 1/2,\mp1/2)$.

The model II corresponds to the two-fold degenerate valence band states hosting heavy holes that are characterized by the angular momentum components $J_z=\pm 3/2$. Such a band structure is typical for self-assembled quantum dots based on III-V and II-VI semiconductors 
\cite{bimbergQuantumDotHeterostructures1999,ivchenko225,efrosNanocrystalQuantumDots2021} and can also be applied to the intravalley states in quantum dots based on transition metal dichalcogenides  \cite{luoQuantumDotBehavior2017,zhangElectrotunableArtificialMolecules2017,pisoniGatetunableQuantumDot2018}. In this case the Bloch basis functions of the valence band are chosen as 
\begin{subequations}
\label{hh}
\begin{align}
&|v.b., + 3/2\rangle = -\frac{\mathcal X + \mathrm i \mathcal Y}{\sqrt{2}} \hspace{0.5 mm}| \hspace{-0.5 mm} \uparrow \hspace{0.5 mm} \rangle,\\
&|v.b., - 3/2\rangle = + \frac{\mathcal X - \mathrm i \mathcal Y}{\sqrt{2}} \hspace{0.5 mm}| \hspace{-0.5 mm} \downarrow \hspace{0.5 mm} \rangle.
\end{align}
\end{subequations}
In this model, $\sigma^+$ light drives the transition $-3/2\to -1/2$ where a pair with hole spin $+3/2$ and electron spin $-1/2$ is created, whereas $\sigma^-$ light drives the transition $+3/2\to +1/2$ where a pair with the hole spin $-3/2$ and electron spin $+1/2$ is created. In the model II transitions in the $z$-polarization are forbidden.

The wavefunctions (\ref{Bloch:67c}), (\ref{Bloch:67d}) and (\ref{hh}) are written in the hole representation. In particular, this means that the interband dipole matrix element 
for the transition with excitation of an electron $|\Gamma_6, s_e \rangle$ and a hole $|\Gamma_7, s_h \rangle$ can be written as
\[
M(s_e, s_h; {\bm e})= - e \int \psi_{\Gamma_6, s_e}^{\dag} ({\bm e}\cdot {\bm r}) {\cal K} \psi_{\Gamma_7, s_h} d {\bm r}\:,
\]
where $e < 0$ is the electron charge, $\psi_n$ are the wave functions (\ref{Bloch:67}) and (\ref{hh}), ${\cal K}$ is the time inversion operator $- {\rm i} \sigma_z {\cal K}_0$ with ${\cal K}_0$ being the operation of complex conjugation. One can check that
 \begin{eqnarray} \label{KK0}
&& {\cal K} |\Gamma_7, \pm 1/2 \rangle = \mp  |\Gamma_7, \mp 1/2\rangle\:, \\ &&  {\cal K} |v.b., \pm 3/2\rangle = \mp  |v.b., \mp 3/2 \rangle\:. \nonumber 
\end{eqnarray}

It is instructive to introduce the basic unit vectors of the $\sigma^+$ and $\sigma^-$ circular polarizations as
\begin{subequations}
\label{orts}
\begin{align}
&\hat{e}_+ = - (\hat{e}_x + \mathrm i \hat{e}_ y)/\sqrt{2},\\
&\hat{e}_- = (\hat{e}_ x - \mathrm i \hat{e}_y)/\sqrt{2},
\end{align}
\end{subequations}
respectively, and the interband dipole matrix element $d_{cv}$ as
\begin{eqnarray}
\label{dcv}
&&d_{cv} = \begin{cases}
M\left( \frac12, \frac12; \hat{e}_+ \right), \quad \hspace{2.7 mm}\mbox{model I},\\
M\left(  -\frac12, \frac32; \hat{e}_+ \right),  \quad \mbox{model II},
\end{cases}  \nonumber \\ && \hspace{0.7 cm}= {\rm i} \frac{e \hbar p_{cv}}{m_0 E_g} \begin{cases} \sqrt{2/3} ,\\ - 1.\end{cases}
\end{eqnarray}
Here $\hat{e}_x , \hat{e}_ y$ are the orts directed along the $x,y$ axes, $m_0$ is the free-electron mass, $E_g$ is the band gap, and $p_{cv} = \int \mathcal S \hat{p}_x  \mathcal X d {\bm r}$ is the interband momentum matrix element, $\hat{p}_x = -\mathrm i \hbar \partial/\partial x$, with the integration carried out over the unit-cell volume.

\subsection{Electrons, holes, excitons, and trions in quantum dots}\label{subsec:states}
We assume a strong confinement regime where the size-quantization energies exceed by far the Coulomb interaction energies of the charge carriers which, in their turn, exceed the fine-structure splittings related to the electron-hole exchange interaction~\cite{seminaLocalizedExcitonsTrions2022}.~In this model the envelope functions of the Coulomb complexes, two-particle excitons and three-particle trions, are presented as combinations of products of single-particle envelopes $\varphi_{i_j}(\bm r_j)$ where the subscript $j$ labels the electrons and holes and $i_j$ enumerates the size-quantized states in the quantum dot. We use the atomic-like nomenclature for the size-quantized states denoting as $1s$ the ground subshell, $2s$, $2p$, etc. the excited subshells. The spin-orbit interaction is neglected, hence, the basic states for the excitons can be chosen in the form~\cite{ivchenko225,seminaLocalizedExcitonsTrions2022}
\begin{equation}
\label{exciton}
\Psi_{X}(\bm r_e, \bm r_h;s_e, s_h) = \varphi_{i_e}(\bm r_e) \varphi_{i_h}(\bm r_h) |s_e,s_h\rangle,
\end{equation}
with the two-particle Bloch amplitude $|s_e,s_h\rangle$ depending on the electron $s_e$ and hole $s_h$ spin components.\footnote{We use the electron-hole representation, hence, $s_h=-s_v$ where $s_v$ is the spin component of the removed valence-band electron.} In accordance with the above selection rules the doublet of states with the total spin $s_e + s_h=\pm 1$ is optically active in the corresponding circular polarization. Moreover, for the $\Gamma_6\times \Gamma_7$ exciton, the states with $s_e+s_h=0$ are active in the $z$-polarization. For the heavy-hole excitons, the states with $s_e + s_h = \pm 2$ are dark. 

The trions are composed of two identical charge carriers and an unpaired one. For specificity we consider $n$-type quantum dots with resident electrons. Hence, the $X^-$ trions we are dealing with comprise two electrons and a hole. In the strong confinement limit considered here the ground state trion corresponds to the two-electron spin-singlet state (we omit $\Gamma_6$ for brevity in the notation of the electron Bloch amplitudes and use the subscripts to distinguish the identical charge carriers)~\cite{ivchenko225,seminaLocalizedExcitonsTrions2022} 
\begin{equation}
\label{Bloch:singlet}
|{\rm singlet}\rangle = \frac{1}{\sqrt{2}}\bigl[|1/2\rangle_1 |-1/2\rangle_2 -|-1/2\rangle_1 |1/2\rangle_2 \bigr],
\end{equation} 
and the wavefunction labeled by the unpaired hole spin $s_h$:
\begin{eqnarray}
&& \Psi_{T,\rm s}(\bm r_1, \bm r_2, \bm r_h;s_h) \nonumber \\&& \hspace{1.5 cm}
= \varphi_{1s}(\bm r_1) \varphi_{1s}(\bm r_2) \varphi_{1s_h}(\bm r_h) |{\rm singlet},s_h\rangle,\label{trion:ground}
\end{eqnarray}
where $\bm r_{1,2}$ are the coordinates of the electrons and we use the subscript $h$ to denote the envelope functions of holes. 

The excited trion states can be both singlet and triplet, because the size quantization prevents dissociation of the three particle complex \cite{kavokinFineStructureQuantumdot2003,cortezOpticallyDrivenSpin2002,edigerFineStructureNegatively2007,
bennyExcitationSpectroscopySingle2012,ivchenko302,seminaLocalizedExcitonsTrions2022}. We assume that the resident electron initially occupies the ground level $1s$ of the quantum dot, whereas the electron-hole pair is generated into excited states. These could be $2s$- or $2p$-subshell states $2p_x$, $2p_y$, $2p_z$. In realistic quantum dots the anisotropy of confinement potential allows one to consider these $p$-shell states independently. For specificity we assume that the point-group symmetry of  confinement potential of a quantum dot or nanocrystal contains the mirror planes $\sigma_x$, $\sigma_y$, $\sigma_z$ perpendicular to the particular axes $x$, $y$, and $z$ which are the principal axes of the system (e.g., the quantum dot is a cuboid of size $a \times b \times c$ or a tri-axial ellipsoid). We consider the situation where the second electron occupies one of the $2p$ orbitals (we omit the subscript $x$, $y$ or $z$) and the hole occupies the $2p_h$ orbital of the same symmetry. This model equally applies to the $2s$-excited pair. 

The excited singlet and triplet trion wavefunction are
\begin{subequations}
\label{trion:excited}
\begin{align}
&\Psi_{T,\rm s}(\bm r_1, \bm r_2, \bm r_h;s_h) = |{\rm singlet};s_h\rangle \label{trion:exciteda} \\ & \hspace{2 mm}  \times \frac{\varphi_{1s}(\bm r_1) \varphi_{2p}(\bm r_2) + \varphi_{2p}(\bm r_1) \varphi_{1s}(\bm r_2)}{\sqrt{2}} \varphi_{2 p_h}(\bm r_h), \nonumber \\& \Psi_{T,\rm t}(\bm r_1, \bm r_2, \bm r_h;s_h) = |{\rm triplet}, S_z;s_h\rangle \label{trion:excitedb} \\ & \hspace{2 mm}
\times  \frac{\varphi_{1s}(\bm r_1) \varphi_{2p}(\bm r_2) - \varphi_{2p}(\bm r_1) \varphi_{1s}(\bm r_2)}{\sqrt{2}} \varphi_{2 p_h}(\bm r_h) . \nonumber
\end{align}
\end{subequations}
Here the singlet spin configuration is similar to that of Eq.~(\ref{Bloch:singlet}), and the triplet spin components are given by
\begin{subequations}
\label{Bloch:triplet}
\begin{align}
&|{\rm triplet},S_z=+1\rangle = |1/2\rangle_1 |1/2\rangle_2,\\&|{\rm triplet},S_z=-1\rangle = |-1/2\rangle_1 |-1/2\rangle_2 ,\\
&|{\rm triplet},S_z =0\rangle \nonumber\\
&\hspace{1cm}= \frac{1}{\sqrt{2}}\left[|1/2\rangle_1 |-1/2\rangle_2 +|-1/2\rangle_1 |1/2\rangle_2 \right],
\end{align}
\end{subequations} 
Because of the antisymmetric requirement, the envelope functions in Eqs.~(\ref{trion:exciteda})~and~(\ref{trion:excitedb}) are respectively symmetric and antisymmetric products of single-particle envelopes. The excited trion state is thus characterized by the spin components of electrons and a hole. There are $8$ states in total: 2 degenerate states stemming from the two-electron singlet and 6 degenerate states stemming from the two-electron triplet. We show below that the electron-hole LRE interaction partially removes this degeneracy. 

The energy splitting between the singlet and triplet states is mainly given by the electron-electron Coulomb interaction\footnote{We neglect the fine structure of the two-electron spectrum induced by the spin--orbit interaction  \cite{glazovSpinorbitEffectElectronelectron2009,glazovFineStructureTwoelectron2010}.}
\begin{equation}
\label{EST}
E_{ST} = \frac{e^2}{\varepsilon_0} \int \frac{\varphi_{2p}(\bm r_1)\varphi_{1s}(\bm r_2) \varphi_{1s}(\bm r_1)\varphi_{2p}(\bm r_2)}{|\bm r_1 - \bm r_2|} d\bm r_1 d\bm r_2,
\end{equation}
where $\varepsilon_0$ is the static dielectric constant and the smooth envelopes are chosen real. 

\section{Electron-hole exchange interaction}\label{sec:HLR}

Our main goal is to analyze the effect of the LRE interaction on the fine structure of the excited trion states. In the past it was also called annihilation interaction \cite{birSymmetryStraininducedEffects1974} and can be considered as a result of the interaction of the electron-hole pair with the induced curl-less (or irrotational) electric field  \cite{pikusExchangeInteractionExcitons1971,ivchenko225}. Thus, we resort to the electrodynamical derivation of the LRE interaction \cite{denisovLongitudinalTransverseExcitons1973,goupalovElectrodynamicalTreatmentElectronhole2003}, the same results can be naturally obtained quantum mechanically by decomposing the electron-hole interaction into the direct and exchange parts~\cite{pikusExchangeInteractionExcitons1971,glazovExcitonFineStructure2014}.

Neglecting the retardation we recall that the curl-less field $\bm E(\bm r)$ induced by the polarization $\bm P(\bm r)$ of the medium satisfies the first Maxwell's equation
\begin{equation}
\label{maxwell:D}
\divv{(\varepsilon_b \bm E + 4\pi \bm P)} =0,
\end{equation}
where $\varepsilon_b$ is the background dielectric constant assumed for simplicity to be the same in the quantum dot and in the surrounding matrix. It is straightforward to check that
\begin{equation}
\label{solution}
E_{\alpha,\bm k} = - \frac{4\pi k_\alpha k_\beta}{\varepsilon_b k^2} P_{\beta, \bm k},
\end{equation}
where $k^2 = k_x^2 + k_y^2 + k_z^2$, $\alpha, \beta$ denote the Cartesian components $x,y,z$ or their circularly polarized combinations $\hat{e}_+, \hat{e}_z, \hat{e}_-$, and the Fourier components are defined in the standard way, $\bm E_{\bm k} = \int  \exp{(-\mathrm i \bm k\bm r)} \bm E(\bm r) d\bm r$, $\bm P_{\bm k} = \int  \exp{(-\mathrm i \bm k\bm r)} \bm P(\bm r) d\bm r$. The electric field $\bm E$ in its turn affects the polarization, producing the energy shift of (with the normalization volume set to unity)
\begin{equation}
\label{energy:classical}
-\int\bm P(\bm r) \cdot \bm E(\bm r)d \bm r = \sum_{\bm k}\frac{4\pi k_\alpha k_\beta}{\varepsilon_b k^2} P_{\alpha,\bm k}^* P_{\beta,\bm k}.
\end{equation}

Let $d_{\alpha;s_e,s_h}$ be the interband matrix element of the $\alpha$ component of the dipole-moment operator calculated between the ground state of the crystal and the electron-hole pair state with electron spin component $s_e$ and hole spin component $s_h$. In the electron-electron picture it corresponds to the $\alpha$-component of the interband dipole matrix element. Expressing the exciton polarization in the second quantization approach via the quantum-field operators of the electrons and holes ($a_{s_e,\bm k}$ and $b_{s_h,\bm k'}$ standing for the annihilation operators, and $a_{s_e,\bm k}^\dag$ and $b_{s_h,\bm k'}^\dag$ standing for the creation operators) we obtain from Eq.~\eqref{energy:classical}
\begin{eqnarray}
\label{H:exch}
&&\hat{\mathcal H}_{\rm exch} = \frac{4\pi}{\varepsilon_b} 
 \sum_{\bm k, \bm q, \bm q'} \frac{k_\alpha k_\beta}{k^2} d_{\alpha;s_e,s_h} d^*_{\beta;s_e',s_h'}\\ &&
 \times a^\dag_{s_e,\bm k/2 + \bm q}b^\dag_{s_h,\bm k/2 - \bm q}b_{s_h',\bm k/2 -\bm q'} a_{s_e',\bm k/2 + \bm q'}. \nonumber
\end{eqnarray}
 Here and below summation over all repeated indices (both spin and Cartesian ones) is implied. The physical meaning of Eq.~\eqref{H:exch} is clear: the electron-hole pair is virtually annihilated and created, as a result the energy shift and coupling of different states occur~\cite{rashbaEffectResonanceExcitation1959,pikusExchangeInteractionExcitons1971,
 denisovLongitudinalTransverseExcitons1973}. The electrodynamic derivation of Eq.~\eqref{H:exch} is fully equivalent to the quantum-mechanical derivation presented for completeness in Appendix~\ref{app:exch}. The latter method unequivocally demonstrates the exchange Coulomb interaction nature of Eq.~\eqref{H:exch}.
 
For the simple $\Gamma_6 \times \Gamma_7$ bands, in accordance with the selection rules formulated above and the relation (\ref{KK0}), we obtain
\begin{subequations}
\label{d:defs}
\begin{eqnarray}
\label{d:simple}
&& d_{\alpha;s_e,s_h} = d^{(1)}_{cv} \biggl[  \delta_{\alpha,\sigma_+} \delta_{s_e,+\frac12} \delta_{s_h,+\frac12} \\ && \hspace{1.2 cm}+\ \delta_{\alpha,\sigma_-} \delta_{s_e,-\frac12}\delta_{s_h,-\frac12} \nonumber \\ && \left.
+\ \frac{\delta_{\alpha,z}}{\sqrt{2}}  \left( \delta_{s_e,+\frac12} \delta_{s_h,-\frac12} +\ \delta_{s_e,-\frac12} \delta_{s_h,+\frac12} \right) \right]. \nonumber 
\end{eqnarray}
In the heavy-hole model~II we simply have
\begin{eqnarray}
\label{d:heavy}
&&d_{\alpha;s_e,s_h} = d^{(2)}_{cv} \left( \delta_{\alpha,\sigma_+} \delta_{s_e,-\frac12}\delta_{s_h,+\frac32} \right.\\ && \left.
\hspace{1 cm}+\  \delta_{\alpha,\sigma_-} \delta_{s_e,+\frac12}\delta_{s_h,-\frac32}\right).\nonumber 
\end{eqnarray}
\end{subequations}
Here $d^{(1)}_{cv}$ and $d^{(2)}_{cv}$ are matrix elements (\ref{dcv}) in the models I and II.

Equations~\eqref{H:exch} and \eqref{d:defs} allow us to evaluate the LRE interaction and corresponding fine structure of the excitons and trions in quantum dots.

For illustration and completeness we briefly recall the fine structure of neutral excitons in quantum dots described by the Hamiltonian~\eqref{H:exch}. We present the exciton wavefunctions in the second-quantized form as
\begin{equation}
\label{exc:2nd}
|{\rm exc}; s_e, s_h\rangle = \sum_{\bm k_e,\bm k_h} F_{\rm exc}(\bm k_e,\bm k_h) a^\dag_{s_e,\bm k_e} b^\dag_{s_h,\bm k_h}|0\rangle,
\end{equation}
where $|0\rangle$ is the ground state of the system, $F_{\rm exc}(\bm k_e,\bm k_h)$ is the Fourier transform of the two-particle exciton envelope function
\begin{equation}
\label{Fexc}
F_{\rm exc}(\bm k_e,\bm k_h) = F_{i_e}(\bm k_e) F_{i_h}(\bm k_h),
\end{equation}
where
$F_i(\bm k) = \int \varphi_i(\bm r)e^{-\mathrm i \bm k\bm r} d\bm r$. Using the representation  \eqref{exc:2nd} we get for the LRE interaction matrix elements of the Hamiltonian \eqref{H:exch} (cf.~\cite{ivchenko225}):
\begin{eqnarray}
\label{exc:fine}
&&\hspace{1 cm}\langle{\rm exc}; s_e, s_h|\hat{\mathcal H}_{\rm exch}|{\rm exc}; s_e', s_h'\rangle \\ && \hspace{1.5 cm}= \frac{4\pi}{\varepsilon_b} \sum_{\bm k, \bm q, \bm q'} \frac{k_\alpha k_\beta}{k^2} d_{\alpha;s_e,s_h} d^*_{\beta;s_e',s_h'} \nonumber\\ &&
\times
F_{\rm exc}^*\left(\frac{\bm k}{2}+\bm q, \frac{\bm k}{2} - \bm q \right)F_{\rm exc} \left(\frac{\bm k}{2}+\bm q', \frac{\bm k}{2} - \bm q' \right). \nonumber
\end{eqnarray}

\begin{figure*}[t]
\includegraphics[width=0.9\textwidth]{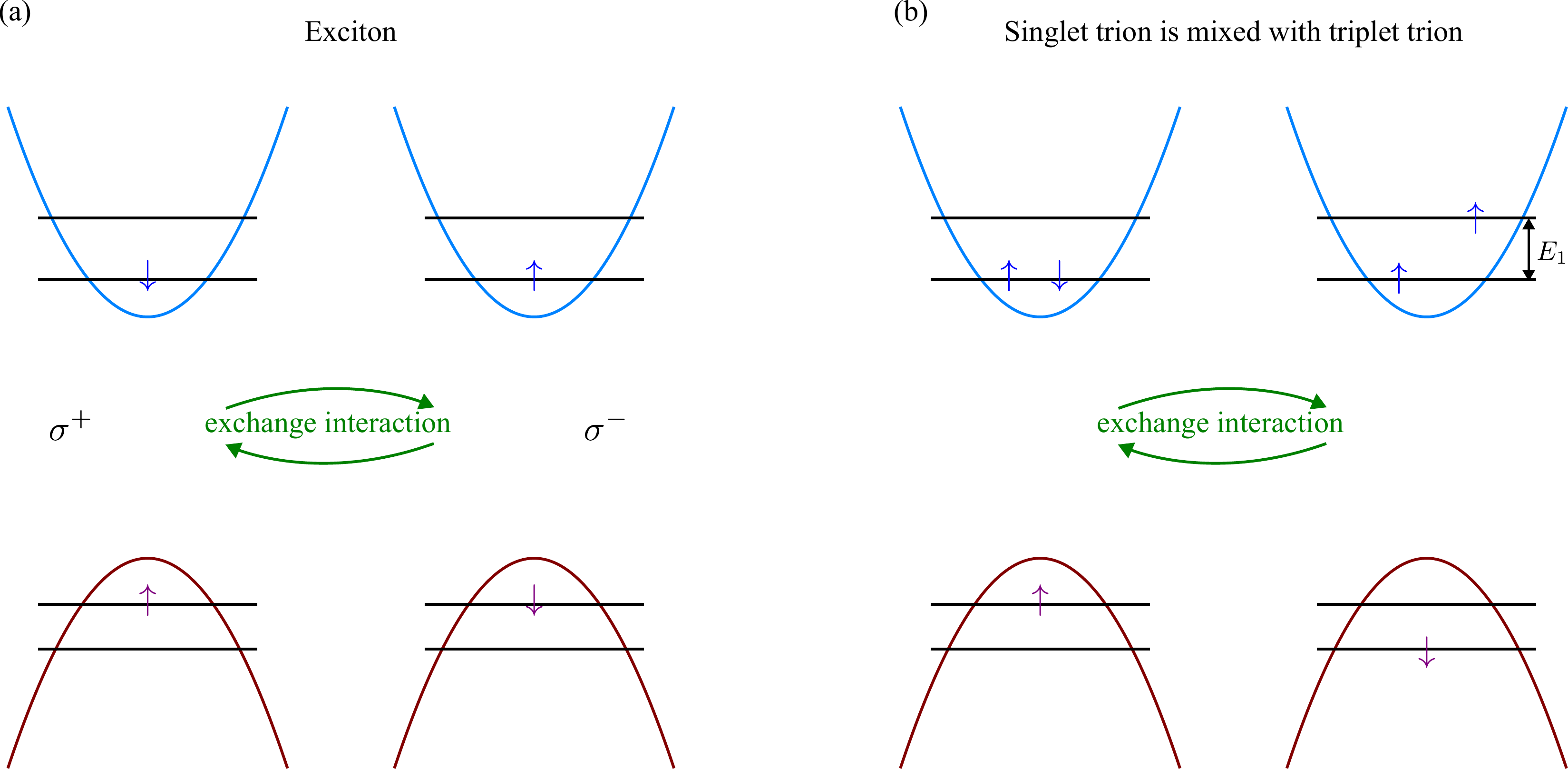}
\caption{Illustration of the effect of LRE interaction on the ground exciton (a) and trion (b) states in the heavy-hole model. For the ground exciton state under anisotropic confinement, the sums $\sum_{\bm k} \left\vert D_{\bm k}^{i_e,i_h}\right\vert^2k_x^2$ and $\sum_{\bm k} \left\vert D_{\bm k}^{i_e,i_h}\right\vert^2k_y^2$ are different, and the $\sigma^+$ and $\sigma^-$ polarized states are mixed. (b) For the ground singlet trion state the mixing is possible only with optically active excited state since two electrons with parallel spins cannot be in the same orbital state.}\label{fig:exc:ill}
\end{figure*}

For the following it is instructive to introduce 
\begin{equation}
\label{D:1s1s}
{D}_{\bm k}^{i_e,i_h} = d_{cv} \int \exp{(- {\rm i} \bm k\bm r)} \varphi_{i_e}(\bm r) \varphi_{i_h}(\bm r) d\bm r,
\end{equation}
and bear in mind that
\begin{eqnarray}
\label{relations}
&&\frac{\left( {\bm k} \cdot \hat{e}^*_+ \right) \left( {\bm k} \cdot \hat{e}_+ \right)}{k^2} = \frac{\left( {\bm k} \cdot \hat{e}^*_- \right) \left( {\bm k} \cdot \hat{e}_- \right)}{k^2} = \frac{k^2_{\parallel}}{2k^2},\\ &&\hspace{1.5 cm}\frac{\left( {\bm k} \cdot \hat{e}^*_+ \right) \left( {\bm k} \cdot \hat{e}_- \right)}{k^2} = - \frac{k_-^2}{2k^2},\nonumber \\ && \hspace{1.5 cm}\frac{\left( {\bm k} \cdot \hat{e}^*_- \right) \left( {\bm k} \cdot \hat{e}_+ \right)}{k^2} = - \frac{k_+^2}{2k^2}, \nonumber
\end{eqnarray}
where, to shorten the notation, we use $k_\parallel^2 = k_x^2 + k_y^2$ and $k_\pm = k_x\pm\mathrm i k_y$. Hence, in the heavy-hole model II, in the basis of the two exciton states active in the $\sigma^+$ and $\sigma^-$ polarizations, the matrix~\eqref{exc:fine} takes a simple form
\begin{equation}
\label{exc:fine:1}
\hat{\cal H}_{\rm exch}^{\rm exc} =
\frac{2\pi}{\varepsilon_b}  \sum_{\bm k} \left\vert D_{\bm k}^{i_e,i_h} \right\vert^2 \frac{1}{k^2}
\left[ \begin{array}{cc}
k_\parallel^2 & -k_-^2\\
-k_+^2 & k_\parallel^2
\end{array} \right].
\end{equation}
The energy is reckoned from the energy of the bright doublet in the absence of LRE interaction and the minus sign in the off-diagonal elements appears because of the sign convention for the unit vector $\hat{e}_+$. The expression for the exciton fine structure splitting agree with those presented in the literature~\cite{ivchenko225,pikusExchangeInteractionExcitons1971,maialleExcitonSpinDynamics1993,gupalovFineStructureLocalized1998,glazovExcitonFineStructure2014}. For the studied case where the shape of quantum dot is characterized by the Cartesian symmetry axes $x, y, z$, the eigenstates of Eq.~\eqref{exc:fine:1} are linearly polarized along these axes: for an anisotropic quantum dot the LRE interaction results in the mixing of $\sigma^+$ and $\sigma^-$ polarized states and formation of linearly polarized sublevels as illustrated in Fig.~\ref{fig:exc:ill}(a).

For simple bands of the model I there are four exciton states. Let, as before, the $x$, $y$, and $z$ axes be the principal axes of the nanocrystal. Then, in agreement with the symmetry and previous works, the exciton fine structure is described by the Hamiltonian
\begin{equation}
\label{exc:fine:1:simple}
{\cal H}_{\rm exch}^{\rm exc} =
\frac{2\pi}{\varepsilon_b}   \sum_{\bm k} \frac{\left\vert D_{\bm k}^{i_e,i_h} \right\vert^2}{k^2} \left[
\begin{array}{cccc}
k_\parallel^2 & -k_-^2 & 0 & 0\\
-k_+^2 & k_\parallel^2 & 0 & 0\\
0 & 0 & k_z^2 & k_z^2 \\
0 & 0 & k_z^2 & k_z^2
\end{array}\right],
\end{equation}
presented here in the basis $|1/2,1/2 \rangle$, $|-1/2,-1/2 \rangle$, $|1/2,-1/2 \rangle$ and $|-1/2,1/2 \rangle$. It should be mentioned that for this choice of axes
\[
\sum_{\bm k} \frac{k_{\pm}^2}{k^2} \left\vert D_{\bm k}^{i_e,i_h} \right\vert^2  = \sum_{\bm k} \frac{k_x^2 - k_y^2}{k^2} \left\vert D_{\bm k}^{i_e,i_h} \right\vert^2 .
\]
In this case the exciton quartet is split into three sublevels polarized along the $x$, $y$, and $z$ axes and one dark sublevel. For a cubic nanocrystal, the three bright states are degenerate and this optically active triplet has the energy 
$$
\frac{4\pi}{3} \sum_{\bm k} \left\vert D_{\bm k}^{i_e,i_h} \right\vert^2 = \frac{4\pi}{3 \varepsilon_b} |d^{(1)}_{cv}\varphi(\bm r = 0)|^2,
$$ 
where $\varphi(\bm r)$ is the envelope function of exciton relative motion. 
Note that, in addition to the LRE interaction, there exists a short-range exchange interaction (also known as isotropic exchange) in the form 
\begin{equation}
\label{eq:short}
{\cal H}_{\rm short} \propto \delta(\bm r_e - \bm r_h),
\end{equation}
which also contributes to the bright-dark exciton splitting~\cite{ivchenko225}. We stress that the mixing of the states active in the opposite circular polarizations being controlled by the off-diagonal matrix elements in Eqs.~\eqref{exc:fine:1} and \eqref{exc:fine:1:simple} and described by the terms $\propto k_\pm^2$  is related to the fact that the curl-less near field produced by the exciton mixes the states active in the opposite circular polarizations despite the fact that the transversal far field emitted by the corresponding exciton is circularly polarized.

Now we come to the main part of our work and address the LRE interaction in trions. The Hamiltonian~\eqref{H:exch} allows one to calculate the fine structure of trion levels in the same way as for the excitons. We present the trion wavefunction in the second quantization approach as
\begin{multline}
\label{tr:2nd}
\hspace{2 cm}|T; s_{1}, s_{2}, s_h\rangle \\
= \sum_{\bm k_1, \bm k_2, \bm k_h} F_{T}^{s_1,s_2}(\bm k_1,\bm k_2,\bm k_h) a^\dag_{s_1,\bm k_1} a^\dag_{s_2,\bm k_2} b^\dag_{s_h,\bm k_h}|0\rangle,
\end{multline}
where $F_{T}^{s_1,s_2}(\bm k_1,\bm k_2,\bm k_h)\equiv -F_{T}^{s_2,s_1}(\bm k_2,\bm k_1,\bm k_h)$ is the Fourier transform of the smooth trion envelope function introduced in Sec.~\ref{subsec:states}. The calculation of the matrix elements is straightforward and gives the following expression for the LRE interaction matrix elements:
\begin{widetext}
\begin{eqnarray}\label{tr:fine}
&&\hspace{0.6 cm}\langle T; s_{1}', s_{2}', s_h'|\hat{\mathcal H}_{\rm exch} |T; s_{1}, s_{2}, s_h\rangle = \frac{4\pi}{\varepsilon_b} \sum_{\bm k, \bm q, \bm q',\bm k_e}
\frac{k_\alpha k_\beta}{k^2} d_{\alpha;s_e,s_h'} d^*_{\beta;s_e',s_h} \\
&&\times\left[
F^{s_1',s_2';*}_{T}(\bm k/2+\bm q,\bm k_e,\bm k/2 - \bm q)
F^{s_1,s_2}_{T}(\bm k/2+\bm q',\bm k_e,\bm k/2 - \bm q')
\delta_{s_e,s_1'}\delta_{s_e',s_1}\delta_{s_2,s_2'}
\right. \nonumber
\\
&&+F^{s_1',s_2';*}_{T}(\bm k_e,\bm k/2+\bm q,\bm k/2 - \bm q)
F^{s_1,s_2}_{T}(\bm k_e,\bm k/2+\bm q',\bm k/2 - \bm q')
\delta_{s_e,s_2'}\delta_{s_e',s_2}\delta_{s_1,s_1'} \nonumber
\\
&&-F^{s_1',s_2';*}_{T}(\bm k_e,\bm k/2+\bm q,\bm k/2 - \bm q)
F^{s_1,s_2}_{T}(\bm k/2+\bm q',\bm k_e,\bm k/2 - \bm q')
\delta_{s_e,s_2'}\delta_{s_e',s_1}\delta_{s_1',s_2} \nonumber
\\ && \left. -F^{s_1',s_2';*}_{T}(\bm k/2+\bm q,\bm k_e,\bm k/2 - \bm q)
F^{s_1,s_2}_{T}(\bm k_e,\bm k/2+\bm q',\bm k/2 - \bm q')
\delta_{s_e,s_1'}\delta_{s_e',s_2}\delta_{s_2',s_1}
\right]. \nonumber
\end{eqnarray}
Introducing
\[
\mathcal F^{s_e,\lambda}_{s_1s_2} (\bm p,\bm k_e,\bm k_h) = \delta_{s_e,s_1}\delta_{\lambda,s_2} F_T^{s_1,s_2}(\bm p,\bm k_e,\bm k_h) - \delta_{s_e,s_2}\delta_{\lambda,s_1} F_T^{s_1,s_2}(\bm k_e,\bm p,\bm k_h),
\]
where $\lambda = \pm 1/2$, we can rewrite Eq.~\eqref{tr:fine} in a more compact form
\begin{eqnarray}
&& \langle T; s_{1}', s_{2}', s_h'|\hat{\mathcal H}_{\rm exch} |T; s_{1}, s_{2}, s_h\rangle
= \frac{4\pi}{\varepsilon_b} \sum_{\bm k,\bm q,\bm q',\bm k_e} \frac{k_\alpha k_\beta}{k^2}
d_{\alpha;s_e,s_h'} d^*_{\beta;s_e',s_h} \nonumber \\ && \hspace{6 mm}\times \sum_\lambda \left[
\mathcal F^{s_e,\lambda}_{s_1's_2'} (\bm k/2+\bm q,\bm k_e,\bm k/2-\bm q) \right]^*
\mathcal F^{s_e',\lambda}_{s_1s_2} (\bm k/2+\bm q',\bm k_e,\bm k/2-\bm q').\nonumber
\end{eqnarray}
\end{widetext}
Physically, the coupling in the form of Eq.~\eqref{tr:fine} results from the virtual annihilation and creation of the electron-hole pair in the trion with the remaining electron acting as a `spectator'. Below we analyze in detail the trion fine structure in the models I and II.
\section{LRE interaction in heavy-hole trions: Model II} \label{subsec:hh:tr}
We start with the technically simpler model of the heavy-hole trion. In this case, an electron-hole pair with given spin components is either optically active in $\sigma^+$ or $\sigma^-$ polarization or dark. It greatly reduces the number of relevant terms in Eq.~\eqref{tr:fine} by virtue of explicit form of the dipole matrix elements, Eq.~\eqref{d:heavy}.
\subsection{Singlet ground state trion in the model II}
For the  singlet state with opposite electron spins and $s_h=\pm 3/2$ the second-quantized trion wavefunction can be written as 
\begin{eqnarray}
\label{tr:2nd:singlet}
&& \hspace{1.5 cm} |{\rm singlet}; 1/2, -1/2, s_h\rangle \\
&&= \sum_{\bm k_1, \bm k_2, \bm k_h} F_{\rm s}(\bm k_1,\bm k_2,\bm k_h) a^\dag_{1/2,\bm k_1} a^\dag_{-1/2,\bm k_2} b^\dag_{s_h,\bm k_h}|0\rangle, \nonumber
\end{eqnarray}
where the singlet envelope is symmetric [cf. Eqs.~\eqref{trion:ground} and \eqref{trion:exciteda}]:
$$F_{\rm s}(\bm k_1,\bm k_2,\bm k_h) =F_{\rm s}(\bm k_2,\bm k_1,\bm k_h).$$
Thus, the singlet trion level is doubly degenerate and lacks a fine structure. Nevertheless, this excitation induces a curl-less electric field and its energy is renormalized because of the LRE interaction as shown below. 

Non-zero matrix elements of the exchange interaction (\ref{H:exch}) form, as expected, a scalar matrix as follows
\begin{eqnarray}
\label{find:singlet:hh}
&&\left\langle {\rm singlet}; \frac12, -\frac12, s_h' \left\vert \hat{\mathcal H}_{\rm exch} \right\vert {\rm singlet}; \frac12, - \frac12, s_h \right\rangle \\ &&
\hspace{1 cm}= \delta_{s_h,s_h'} \frac{2\pi}{\varepsilon_b} |d_{cv}|^2 \sum_{\bm k,\bm q,\bm q',\bm k_e} \frac{k_\parallel^2}{k^2}  \nonumber \\ &&
\times F_{\rm s}^*\left( \frac{\bm k}{2}+\bm q,\bm k_e, \frac{\bm k}{2} - \bm q \right) F_{\rm s} \left(\frac{\bm k}{2}+\bm q',\bm k_e,\frac{\bm k}{2}-\bm q' \right)  \nonumber \\ &&
\hspace{1 cm}=\delta_{s_h,s_h'} \frac{2\pi}{\varepsilon_b} \sum_{\bm k} \left\vert D_{\bm k}^{1s,1s_h}\right\vert^2 \frac{k_\parallel^2}{k^2}, \nonumber
\end{eqnarray}
where the last equality assumes  Eq.~\eqref{trion:ground} for the envelope function. In agreement with the time-reversal symmetry, the half-integer-spin trion state remains two-fold degenerate regardless of specific shape of the confining potential and wavefunctions. Equation~\eqref{find:singlet:hh} can be interpreted as a trion Lamb shift similarly to the light-matter coupling induced renormalization of the exciton resonant frequency~\cite{ivchenko225,renControlBrightDarkExciton2023}. Virtual recombination of the electron-hole pair in the trion can subsequently result in the formation of the singlet trion with the same hole spin component, otherwise two electrons should be in the same-spin triplet state, see Fig.~\ref{fig:exc:ill}(b). Hence, instead of splitting the ground trion state, the LRE interaction mixes it with the excited triplet state. This second-order renormalization is weak due to the large energy separation between  the $1s$ and $2p/2s$ confinement energies. The mixing, however, becomes important for the excited $2p$--$2p$ trion fine structure.

It is worth to mention that, although the LRE interaction does not lead to a splitting of the trion sublevels, this interaction can manifest itself in a pair of two close-lying charged quantum dots with close energy levels: due to the curl-less electric field induced by a trion in one of the quantum dots, the trion excitation can be transferred from one dot to another (the F{\"o}rster effect). Indeed, virtual recombination an electron-hole pair in the first charged dot and its re-excitation in the second charged dot allows for resonant or nearly-resonant trion transfer without  tunneling of charge carriers.

\subsection{Excited $2p$--$2p$ trion states in the model II}
\begin{figure*}[htb]
\includegraphics[width=\textwidth]{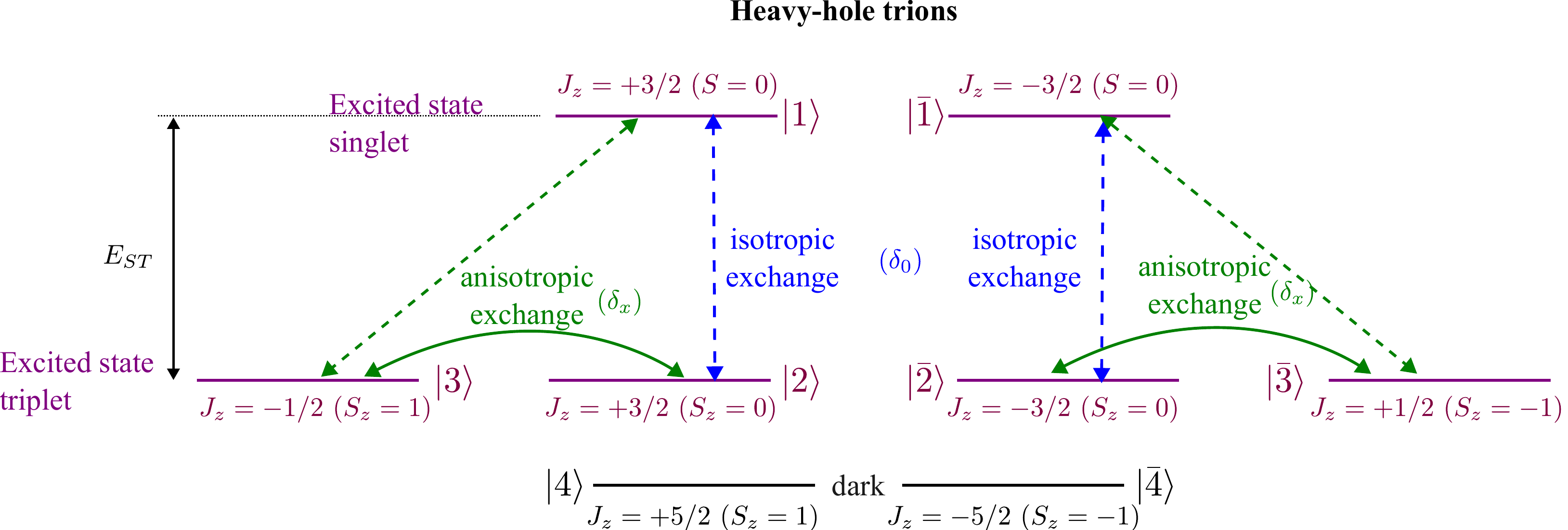}
\caption{Mixing scheme for the states of  heavy hole excited $2p$--$2p$ trions. Left part shows the states in Eqs.~\eqref{pp:states:2nd} and the right part shows their time-reversal partners, Eqs.~\eqref{pp:states:2nd:inv}. The states are labeled by the total spin component $J_z = s_1 + s_2 + s_h$ and the total spin of two electrons. Solid double-sided arrows show triplet-triplet mixing and dashed arrows singlet-triplet mixing. }\label{fig:hh}
\end{figure*}

For the excited trion states we consider both two-electron singlet and triplet, i.e., eight configurations in total. We present explicitly four of them in the second-quantization representation, cf. Eq.~\eqref{trion:excited}:\\
\begin{subequations} \label{pp:states:2nd}
(i) the excited \emph{singlet} state with opposite electron spins ($S=0,S_z=0$), the hole spin
 $s_h=3/2$ and total angular momentum component $J_z = S_z + s_h=+3/2$:
\begin{multline} \label{tr:2nd:singlet:exc} %\label{tr:2nd:singlet:exc}
\left\vert {\rm singlet}^*; \frac12, - \frac12, \frac32 \right\rangle \equiv \left\vert {\rm singlet}^*; J_z= \frac32 \right\rangle \\
=\sum_{\bm k_1, \bm k_2, \bm k_h} \frac{ F_{1s}(\bm k_1)F_{2p}(\bm k_2) + F_{1s}(\bm k_2) F_{2p}(\bm k_1) }{\sqrt{2}}\\ \hspace{1.5 cm}
\times\  F_{2p_h}(\bm k_h)\hspace{0.4 mm}  a^\dag_{1/2,\bm k_1} a^\dag_{-1/2,\bm k_2} b^\dag_{3/2,\bm k_h}|0\rangle; 
\end{multline}
~(ii) the excited \emph{triplet} state with opposite electron spins ($S=1,S_z=0$), the hole spin $s_h=3/2$ and $J_z=+3/2$:
\begin{multline}
\label{tr:2nd:triplet:0:hh}
\left\vert {\rm triplet}^*; \frac12, - \frac12, \frac32 \right\rangle \equiv \left\vert {\rm triplet}^*; J_z= \frac32 \right\rangle \\
=\sum_{\bm k_1, \bm k_2, \bm k_h} \frac{ F_{1s}(\bm k_1)F_{2p}(\bm k_2) - F_{1s}(\bm k_2) F_{2p}(\bm k_1) }{\sqrt{2}}  \\ \hspace{1.5 cm}
\times\  F_{2p_h}(\bm k_h)\hspace{0.4 mm} a^\dag_{1/2,\bm k_1} a^\dag_{-1/2,\bm k_2} b^\dag_{+3/2,\bm k_h}|0\rangle,
\end{multline}
~(iii) the excited \emph{triplet} state with parallel electron spins ($S=1,S_z=+1$), $s_h = - 3/2$ and $J_z=-1/2$:
\begin{multline}
\label{tr:2nd:triplet:minus1:hh}
\left\vert{\rm triplet}^*; \frac12, \frac12, - \frac32 \right\rangle \equiv \left\vert {\rm triplet}^*; J_z=- \frac12 \right\rangle\\
= \sum_{\bm k_1, \bm k_2, \bm k_h} \frac{F_{1s}(\bm k_1)F_{2p}(\bm k_2) - F_{1s}(\bm k_2)F_{2p}(\bm k_1)}{2} \\  \hspace{1.5 cm}
\times\  F_{2p_h}(\bm k_h)\hspace{0.4 mm} a^\dag_{+1/2,\bm k_1} a^\dag_{+1/2,\bm k_2} b^\dag_{-3/2,\bm k_h}|0\rangle  \\
\equiv \sum_{\bm k_1, \bm k_2, \bm k_h}  F_{1s}(\bm k_1)F_{2p}(\bm k_2)F_{2p_h}(\bm k_h)  \\  \hspace{1.5 cm} \times\  a^\dag_{+1/2,\bm k_1} a^\dag_{+1/2,\bm k_2} b^\dag_{-3/2,\bm k_h}|0\rangle, 
\end{multline}
~(iv) the excited \emph{triplet} state  with parallel electron spins\\ ($S=1,S_z=+1$), $s_h = 3/2$ and $J_z=5/2$:
\begin{multline}
\label{tr:2nd:triplet:dark}
\left\vert{\rm triplet}^*; \frac12, \frac12, \frac32 \right\rangle \equiv \left\vert {\rm triplet}^*; J_z=\frac52 \right\rangle\\
 \equiv \sum_{\bm k_1, \bm k_2, \bm k_h}  F_{1s}(\bm k_1)F_{2p}(\bm k_2)F_{2p_h}(\bm k_h) \\ \hspace{1.5 cm} \times\  a^\dag_{+1/2,\bm k_1} a^\dag_{+1/2,\bm k_2} b^\dag_{+3/2,\bm k_h}|0\rangle.
\end{multline}
\end{subequations}
For brevity these four states are labeled as $| n \rangle$ ($n = 1$-$4$), see Fig.~\ref{fig:hh}.
Four more states are obtained from~\eqref{tr:2nd:singlet:exc}, \eqref{tr:2nd:triplet:0:hh}, \eqref{tr:2nd:triplet:minus1:hh} and \eqref{tr:2nd:triplet:dark} by the time reversal operation, namely,
\begin{subequations}
\label{pp:states:2nd:inv}
\begin{align}
&|\bar 1\rangle = |{\rm singlet}^*; 1/2, -1/2, -3/2\rangle, \label{2nd:inv:1}\\
&|\bar 2\rangle = |{\rm triplet}^*; 1/2, -1/2, -3/2\rangle, \label{2nd:inv:2}\\
&|\bar 3\rangle =  |{\rm triplet}^*; -1/2, -1/2,  +3/2\rangle, \label{2nd:inv:3}\\
&|\bar 4\rangle =  |{\rm triplet}^*; -1/2, -1/2, -3/2\rangle.\label{2nd:inv:4}
\end{align}
\end{subequations}

According to the selection rules related to the heavy-hole transitions, see Sec.~\ref{subsec:bands} for details, the first two states in the set~\eqref{pp:states:2nd} are active in the $\sigma^+$ polarization, whereas the third one is active in the $\sigma^-$ polarization (after recombination these three states leave behind an electron with $s_z=+1/2$), and the state~\eqref{tr:2nd:triplet:dark} is dark. Similarly, the first three states in the time-reversed set \eqref{pp:states:2nd:inv} are optically active (first two in the  $\sigma^-$, and the third one in the $\sigma^+$ circular polarization leaving electron with $s_z=-1/2$ after the recombination), while the last sublevel is dark. The states are depicted in Fig.~\ref{fig:hh}.

One can readily check that the states $|1\rangle, |2\rangle, |3\rangle$ [Eqs. \eqref{tr:2nd:singlet:exc}, \eqref{tr:2nd:triplet:0:hh}, and \eqref{tr:2nd:triplet:minus1:hh}] are mixed by the LRE interaction Hamiltonian~\eqref{H:exch}. The matrix elements of this mixing in the basis \eqref{pp:states:2nd}  are
\begin{eqnarray}
\label{H:LT:pp}
&& \hspace{2 cm}{\cal H}_{\rm long} = \frac{\pi}{\varepsilon_b} \sum_{\bm k} \frac{1}{k^2}   \\
&& \times \left[
 \begin{array}{ccc}
{D}_{2,\bm k}  k_\parallel^2 &
\tilde{D}_{2,\bm k} k_\parallel^2 &
-\sqrt{2} \tilde{D}_{2,\bm k} k_-^2\\
\tilde{D}_{2,\bm k} k_\parallel^2 &
{D}_{2,\bm k}  k_\parallel^2 &
-\sqrt{2}{D}_{2,\bm k}  k_-^2\\
-\sqrt{2}\tilde{D}_{2,\bm k} k_+^2 &
-\sqrt{2}{D}_{2,\bm k} k_+^2&
2 { D}_{2,\bm k}  k_\parallel^2
\end{array} \right], \nonumber
\end{eqnarray}
where 
\begin{subequations}
\label{Dddd}
\begin{align}
{D}_{2,\bm k} = |{D}_{\bm k}^{1s,2p_h}|^2 +|{D}_{\bm k}^{2p,2p_h}|^2, \\ 
\tilde{D}_{2,\bm k} = |{D}_{\bm k}^{1s,2p_h}|^2 - |{D}_{\bm k}^{2p,2p_h}|^2,
\end{align}
\end{subequations}
and the Fourier components ${D}_{\bm k}^{i,j}$ are introduced by Eq.~\eqref{D:1s1s}.\footnote{Note that the contributions $\propto |{D}_{\bm k}^{1s,2p}|^2$ appeared in the study of the fine structure of the $s$-$p$ excited exciton in Ref.~\cite{ivchenko231}, and the contributions $\propto |{D}_{\bm k}^{2p,2p}|^2$ for the $p$--$p$ exciton appeared in Ref.~\cite{ivchenko232}.} 

Similarly, the states \eqref{2nd:inv:1}, \eqref{2nd:inv:2} and \eqref{2nd:inv:3} are mixed by the same interaction~\eqref{H:exch}. The effective exchange Hamiltonian for this triplet is given by the complex conjugate of Eq.~\eqref{H:LT:pp}. In accordance with the time reversal symmetry there is no mixing between the sets~\eqref{pp:states:2nd} and~\eqref{pp:states:2nd:inv}.\footnote{Microscopically, it is because to go from the set \eqref{pp:states:2nd} to the set~\eqref{pp:states:2nd:inv} the single electron-hole pair spin flip is insufficient: either one charge carrier should flip its spin, or two electrons, or all three particles.} Hence, it suffices to consider the set of three states \eqref{tr:2nd:singlet:exc}, \eqref{tr:2nd:triplet:0:hh}, and \eqref{tr:2nd:triplet:minus1:hh} separately.
The full Hamiltonian of this set also contains three terms
\begin{equation} \label{3terms}
{\cal H}_{\rm exch} = {\cal H}_{ST} + {\cal H}_{\rm long} + {\cal H}_{\rm short}\:,
\end{equation}
where ${\cal H}_{\rm long}$ is given by Eq.~(\ref{H:LT:pp}), the diagonal matrix
\begin{equation}
\label{EST:mat}
{\cal H}_{ST}= \left[ \begin{array}{ccc}
E_{ST} & 0 & 0\\
0 & 0 & 0 \\
0 & 0 & 0
\end{array} \right],
\end{equation}
describes the singlet--triplet splitting (the largest term), see Eq.~\eqref{EST}, and the last matrix 
\begin{equation}
\label{iso:mat:hh}
{\cal H}_{\rm short}= \left[ \begin{array}{ccc}
\delta_{\rm iso} & \tilde \delta_{\rm iso} & 0\\
\tilde \delta_{\rm iso} & \delta_{\rm iso} & 0 \\
0 & 0 & 2\delta_{\rm iso}
\end{array} \right],
\end{equation}
describes the isotropic short-range exchange interaction, Eq.~\eqref{eq:short}, that mixes the states of the same symmetry and is characterized by two parameters, 
\begin{subequations}
\label{delta:iso}
\begin{align}
\delta_{\rm iso}= C \int d\bm r |\varphi_{2p_h}(\bm r)|^2 (|\varphi_{1s}(\bm r)|^2 + |\varphi_{2p}(\bm r)|^2),\\
\tilde \delta_{\rm iso}  = C \int d\bm r |\varphi_{2p_h}(\bm r)|^2 (|\varphi_{1s}(\bm r)|^2 - |\varphi_{2p}(\bm r)|^2),
\end{align}
with $C$ being a coefficient related to the short-range exchange interaction.
\end{subequations}

\begin{figure*}
\includegraphics[width=\textwidth]{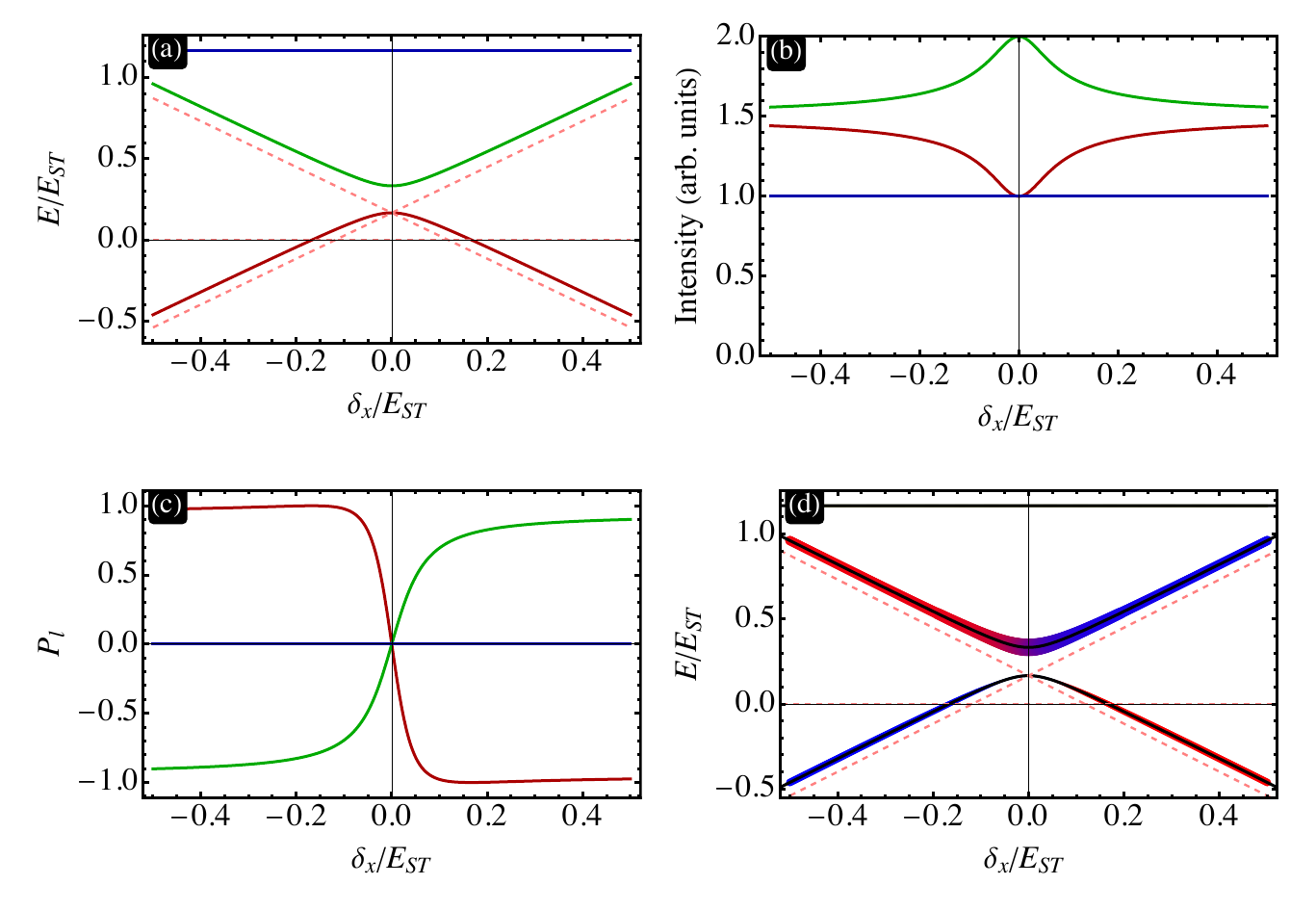}
\caption{Energy spectrum of excited heavy-hole trion states calculated within the \textbf{model II.1}, Eq.~\eqref{model1}, with $\delta_0=E_{ST}/6$. Panel (a) shows the energies, panel (b) shows the intensities of the emission (oscillator strengths), and panel (c) shows the degrees of linear polarization. Blue lines represent the singlet state, red and green lines show the combinations of triplet states $J_z=3/2,-1/2$ (or $J_z=- 3/2, 1/2$) mixed by the Hamiltonian~\eqref{H23}. Panel (d) shows the emission intensity (line thickness) and linear polarization in the $x$-$y$ axes frame (color from blue corresponding to the $x$-polarization to red corresponding to $y$-polarization) superimposed on the energy spectrum. Thin red dashed lines in (a) and (d) indicate asymptotics $E=0$ and $E_\pm =\delta_0 \pm \sqrt{2}\delta_x$. Energy is reckoned from the triplet trion state at $\delta_0=0$.}\label{fig:model1}
\end{figure*}

Figure~\ref{fig:hh} shows the scheme of levels and coupling between them for  the excited trion states based on the analysis above. To gain further insight it is instructive to consider two simplified models. In the \textbf{model II.1}, we assume that the overlap integrals in Eqs.~\eqref{Dddd} and \eqref{delta:iso}  between the $2p_h$ hole state and $1s$ electron state are the same as for the $2p_h$ hole and $2p$ electron such that 
\begin{equation}
\label{model1}
\tilde{D}_{2,\bm k}=0,\quad \tilde \delta_{\rm iso}=0.
\end{equation} 
In this case the excited singlet state becomes decoupled from excited triplet states (the contribution of dashed arrows vanishes). Such model allows us to minimize the number of involved states. The effective Hamiltonian in the basis of the states $|2\rangle= | {\rm triplet}^*; J_z=+3/2\rangle$ and $|3\rangle = | {\rm triplet}^*; J_z=-1/2\rangle$ takes a simple form
\begin{equation}
\label{H23}
{\cal H}_{23} = \left[ \begin{array}{cc}
\delta_0 & - \sqrt{2}\delta_x \\
- \sqrt{2}\delta_x & 2\delta_0
\end{array} \right],
\end{equation}
where 
\begin{subequations}
\label{delta0x}
\begin{align}
&\delta_0 = \delta_{\rm iso} + \frac{\pi}{\varepsilon_b} \sum_{\bm k} {D}_{2,\bm k} \frac{k_\parallel^2}{k^2},\\
&\delta_x = \frac{\pi}{\varepsilon_b} \sum_{\bm k}{D}_{2,\bm k} \frac{k_x^2 - k_y^2}{k^2},
\end{align}
\end{subequations}
and we assumed that $x$ and $y$ are the in-plane principal axes of the structure. Such coupling corresponds to solid double-headed arrows in Fig.~\ref{fig:hh}. The parameter $\delta_x$ is related to the anisotropy of the system and it is solely controlled by the LRE interaction. Similar Hamiltonian governs the fine structure of the time-reversed pair~\eqref{2nd:inv:2} and \eqref{2nd:inv:3}. Hence, the eigenspectrum of excited trions in this model consists of $4$ doubly degenerate states: dark triplets with $J_z=\pm 5/2$ (their energy is $0$ in our frame of reference), singlets with $J_z =\pm 3/2$ (one is active in the $\sigma^+$ polarization and the other is $\sigma^-$ polarized) with the energy $E_{ST} + \delta_0$, and the combinations of the states \eqref{tr:2nd:triplet:0:hh} and \eqref{tr:2nd:triplet:minus1:hh} with $J_z=3/2, -1/2$, and the states \eqref{2nd:inv:2}, \eqref{2nd:inv:3} with $J_z =-3/2,1/2$ which are, generally, mixed by the anisotropic part of the LRE interaction and linearly polarized along the principal axes of the quantum dot.  

The dimensionless energy of the excited trion states, $E/E_{ST}$, intensity of their emission, or oscillator strength, $I$, and emission linear polarization $P_l$ in the principal axes of the structure are shown in Fig.~\ref{fig:model1}. In this model the excited singlet (blue line) is decoupled from the two optically active triplet states. The Kramers-degenerate dark doublets $|4\rangle$ and $|\bar 4\rangle$ have zero energy unaffected by the non-zero value of $\delta_x$. At zero $\delta_x$ the states $|2\rangle$ and $|3\rangle$ (or $|\bar 2\rangle$ and $|\bar 3\rangle$) are split by $\delta_0$; for non-zero $\delta_x$ they become linearly polarized. An anticrossing of these states with the increasing $\delta_x$ is clearly visible. It is controlled by an interplay between the anisotropic and isotropic splittings; 

\begin{figure}
\includegraphics[width=\linewidth]{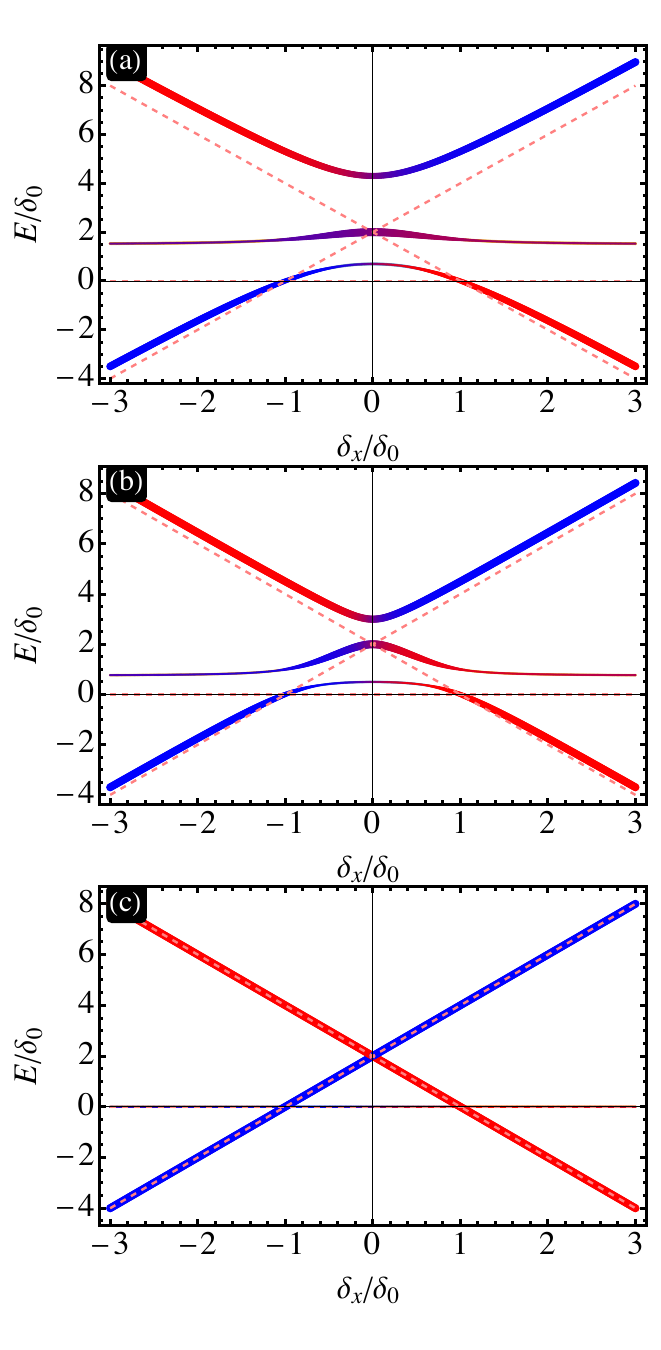}
\caption{Energy spectrum of excited trion states calculated within the \textbf{model II.2}, Eq.~\eqref{model2}. Panels (a)--(c) are calculated for $E_{ST}/\delta_0=3$ (a), $1.5$ (b), and $0$ (c) and show the emission intensity (line thickness) and linear polarization in the $x$--$y$ axes frame (color from blue corresponding to the $x$-polarization to red corresponding to the $y$-polarization) superimposed on the energy spectrum. Thin red dashed lines show asymptotics $E=0$ (dark states) and $E_\pm$ in Eq.~\eqref{limit:doublets}. Energy is reckoned from the dark triplet states $|4\rangle$, $|\bar 4\rangle$.}\label{fig:model2}
\end{figure}

Within the \textbf{model II.2} we assume that  overlap between the $2p_h$ hole state and $1s$ electron state is formally zero. In this situation 
\begin{equation}
\label{model2}
{D}_{\bm k}^{1s,2p_h}=0,~~\tilde{{D}}_{2,\bm k} = - {D}_{2,\bm k},~~\tilde \delta_{\rm iso} = - \delta_{\rm iso}.
\end{equation} 
This model allows us to minimize the number of parameters keeping the singlet-triplet mixing and parametrize the $3\times 3$ Hamiltonian of the system via the same two quantities $\delta_0$ and $\delta_x$ introduced in Eqs.~\eqref{delta0x}. The model explicitly demonstrates the coupling between the singlet and triplet states, see Fig.~\ref{fig:hh} where both solid and dashed double-headed arrows demonstrate the corresponding mixing. Figure~\ref{fig:model2} shows the evolution of the spectrum and linear polarization of states as a function of anisotropic LRE splitting $\delta_x$ for three ratios of the singlet--triplet splitting to the total isotropic splitting, $E_{ST}/\delta_0$. Here, the mixing of all  optically active states (both singlet and triplet) and their anticrossings with increasing $\delta_x$ are seen. For sufficiently large $|\delta_x| > \delta_0$ one of the states becomes optically inactive, while remaining two acquire linear polarization in the $x$--$y$ axes frame: one along the $x$-axis and the other one along the $y$-axis. Note that the energies and polarization properties of the Kramers-partner states ($|\bar 1\rangle \ldots |\bar 4\rangle$) are exactly the same.

It is interesting to trace the transition to the limit of an exciton, when the resident (in this situation, $1s$) electron is moved far away from the active $2p$--$2p_h$ pair. It is readily done within the model II.2 by letting $E_{ST}$ in Eqs.~\eqref{EST} and~\eqref{EST:mat} approach zero. Then one doubly degenerate dark state remains (a superposition of the singlet and triplet states with $S_z=0$), and two pairs of doublets appear with the splitting equal to the anisotropic exciton splitting, see Fig.~\ref{fig:model2}(c):
\begin{equation}
\label{limit:doublets}
E_{\pm} = 2 (\delta_0 \pm \delta_x).
\end{equation}
Here, as before, the principal axes of the system are assumed to be directed along the $x$ and $y$ axes. Each of the levels $E_\pm$ is doubly degenerate in the spin of the remote $1s$ electron. Equation~\eqref{limit:doublets} allows us to clarify the physical meaning of the parameters $\delta_0$ and $\delta_x$: $2\delta_0$ describes the isotropic bright-dark splitting of the $2p$--$2p_h$ exciton and $2\delta_x$ describes the anisotropic splitting of its radiative doublet. A transformation of spectra with decrease of $E_{ST}$ at a fixed $\delta_0$ as a function of $\delta_x$ is shown in Figs.~\ref{fig:model2}(a--c).

\section{Trions in nanocrystals with simple bands: Model I} \label{simple bands}

To evaluate the LRE interaction for the trions in quantum dots with the simple band structure $\Gamma_6 \times \Gamma_7$ we need to use the dipole matrix elements in Eq.~\eqref{d:simple} and take also into account the $z$-polarized transition. 

\subsection{Singlet trion in the ground state in the model I}

As in the case of trions with a heavy hole, if the coupling to excited states is neglected, there is only a diagonal matrix element responsible for the ``Lamb'' shift of a singlet trion:
\begin{multline}
\label{find:singlet:simple}
\langle {\rm singlet}; 1/2, -1/2, s_h'|\hat{\cal H}_{\rm exch} |{\rm singlet}; 1/2, -1/2, s_h\rangle \\
= \delta_{s_h,s_h'} \frac{2\pi}{\varepsilon_b} |d_{cv}|^2  \\ 
\times \sum_{\bm k,\bm q,\bm q',\bm k_e}  F_{\rm s}^*(\bm k/2+\bm q,\bm k_e,\bm k/2 - \bm q) F_{\rm s} (\bm k/2+\bm q',\bm k_e,\bm k/2 - \bm q')\\ 
=\delta_{s_h,s_h'} \frac{2\pi}{\varepsilon_b} \sum_{\bm k} |{D}_{\bm k}^{1s,1s_h}|^2. \hspace{0.8 cm}
\end{multline}
The comparison of Eq.~\eqref{find:singlet:simple} with its counterpart for heavy holes, Eq.~\eqref{find:singlet:hh}, shows that the only important difference is the factor $ (k_\parallel^2+k_z^2)/k^2 = 1$ in the model I as compared to the factor $k_\parallel^2/k^2$ for the model II. This is because, for the $\Gamma_6 \times \Gamma_7$ bands, the singlet trion is active both in the in-plane and out-of-plane polarizations. Clearly, the Kramers degeneracy of the two states with $s_h=\pm 1/2$ cannot be broken by anisotropy of the confinement.

\subsection{Excited $2p$--$2p$ trion state in the model I}

The second-quantized wavefunctions of the excited $2p$--$2p$ trions can be constructed in a way, similar to that of the heavy-hole trions. It is also convenient to separate the full set of $8$ states into two sets of $4$ states each related by the time reversal symmetry. The first set can be presented as follows:\\
\begin{subequations}
\label{pp:states:2nd:simple}
(i) The excited singlet state with opposite electron spins ($S=0,S_z=0$) and $s_h = 1/2$:
\begin{multline}
\label{tr:2nd:singlet:exc:simple}
\hspace{1.5 cm}|1\rangle = |{\rm singlet}^*; 1/2, -1/2, 1/2\rangle\\
= \sum_{\bm k_1, \bm k_2, \bm k_h} \frac{F_{1s}(\bm k_1)F_{2p}(\bm k_2) + F_{1s}(\bm k_2)F_{2p}(\bm k_1)}{\sqrt{2}} \\
\times F_{2p_h}(\bm k_h) a^\dag_{1/2,\bm k_1} a^\dag_{-1/2,\bm k_2} b^\dag_{1/2,\bm k_h}|0\rangle,
\end{multline}
~(ii) the excited triplet state with opposite electron spins ($S=1,S_z=0$) and $s_h = 1/2$:
\begin{multline}
\label{tr:2nd:triplet:0:simple} \hspace{1.5 cm}
|2\rangle = |{\rm triplet}^*; 1/2, -1/2, 1/2\rangle  \\
= \sum_{\bm k_1, \bm k_2, \bm k_h} \frac{F_{1s}(\bm k_1)F_{2p}(\bm k_2) - F_{1s}(\bm k_2)F_{2p}(\bm k_1)}{\sqrt{2}}\\
 \times F_{2p_h}(\bm k_h) a^\dag_{1/2,\bm k_1} a^\dag_{-1/2,\bm k_2} b^\dag_{1/2,\bm k_h}|0\rangle,
\end{multline}
~(iii) the excited triplet state with parallel electron spins ($S=1,S_z=-1$) and $s_h = - 1/2$:
\begin{multline}
\label{tr:2nd:triplet:1:simple}
\hspace{1.5 cm}|3\rangle = |{\rm triplet}^*; -1/2, -1/2, -1/2\rangle \\
= \sum_{\bm k_1, \bm k_2, \bm k_h}  F_{1s}(\bm k_1)F_{2p}(\bm k_2)\\
\times  F_{2p_h}(\bm k_h) a^\dag_{-1/2,\bm k_1} a^\dag_{-1/2,\bm k_2} b^\dag_{-1/2,\bm k_h}|0\rangle ,
\end{multline}
~(iv) the excited triplet state with parallel electron spins ($S=1,S_z=+1$) and the opposite hole spin $s_h = - 1/2$:
\begin{multline}
\label{tr:2nd:triplet:111:simple}
\hspace{1.5 cm}|4\rangle = |{\rm triplet}^*; 1/2, 1/2, -1/2\rangle  \\
 = \sum_{\bm k_1, \bm k_2, \bm k_h}  F_{1s}(\bm k_1)F_{2p}(\bm k_2)\\
 \times F_{2p_h}(\bm k_h) a^\dag_{+1/2,\bm k_1} a^\dag_{+1/2,\bm k_2} b^\dag_{-1/2,\bm k_h}|0\rangle.
\end{multline}
\end{subequations}
The second set consists of their time-reversal partners $|\bar 1\rangle  \ldots |\bar 4\rangle$.

It follows from the selection rules for the simple $\Gamma_6 \times \Gamma_7$ bands (Sec.~\ref{subsec:bands}) that the two states $|1\rangle$, $|2\rangle$ are active respectively in the $\sigma^+$ polarization (after recombination the $s_z=-1/2$ electron remains) and $z$ polarization (after recombination the $s_z=+1/2$ electron remains), the third state $|3\rangle$ is active in the $\sigma^-$ polarization (leaving behind the $s_z=-1/2$ electron after recombination), and the fourth state is active in the $z$ polarization (leaving the $s_z=+1/2$ electron after recombination). These states are mixed by Hamiltonian~\eqref{H:exch}. Similarly, the Kramers partners of the states are active in the $\sigma^-$ ($|\bar{1}\rangle$), $\sigma^+$ ($|\bar{3}\rangle$) and $z$ ($|\bar{2} \rangle$, $|\bar{4} \rangle$) polarizations, and they are also mixed by the LRE interaction.  In accordance with the Kramers theorem the two sets are not mixed by the LRE interaction similarly to the heavy-hole case.

Similarly to the heavy-hole case, in order to establish the excited trion fine structure it suffices to analyze one set of states, e.g., $|1\rangle \ldots |4\rangle$. The Hamiltonian of the LRE-induced mixing in this basis of states can be written as 
\begin{widetext}
\begin{equation}
\label{H:LT:pp:simple}
{\cal H}_{\rm exch} = \frac{\pi}{\varepsilon_b} \sum_{\bm k} \left[
 \begin{array}{cccc}
{D}_{2,\bm k} &
\tilde{D}_{2,\bm k}\frac{k_\parallel^2 - k_z^2}{k^2} &
-\sqrt{2}\tilde{D}_{2,\bm k} \frac{k_x^2-k_y^2}{k^2} & \sqrt{2}\tilde{D}_{2,\bm k} \frac{k_z^2}{k^2} \\
\tilde{D}_{2,\bm k}\frac{k_\parallel^2 - k_z^2}{k^2} &
{D}_{2,\bm k} &
-\sqrt{2}{D}_{2,\bm k} \frac{k_x^2-k_y^2}{k^2} & -\sqrt{2}{D}_{2,\bm k} \frac{k_z^2}{k^2}\\
-\sqrt{2}\tilde{D}_{2,\bm k} \frac{k_x^2-k_y^2}{k^2} &
-\sqrt{2}{D}_{2,\bm k} \frac{k_x^2-k_y^2}{k^2}&
2{D}_{2,\bm k}\frac{k_\parallel^2}{k^2} & 0 \\
\sqrt{2}\tilde{D}_{2,\bm k} \frac{k_z^2}{k^2} &  -\sqrt{2}{D}_{2,\bm k} \frac{k_z^2}{k^2} & 0 &  2{D}_{2,\bm k} \frac{k_z^2}{k^2}
\end{array} \right],
\end{equation}
\end{widetext}
where an existence of principal axes $x$, $y$, $z$ is assumed in the quantum dot, and the quantities ${D}_{2, \bm k}$, $\tilde{D}_{2,\bm k}$ are introduced in Eq.~\eqref{Dddd}.\footnote{In the case where the nanocrystal shape is characterized by three perpendicular mirror planes (see Sec.~\ref{subsec:states}), two symmetric quadratic forms $D_{\alpha\beta} = \sum_{\bm k} {\cal  D}_{2,\bm k} k_\alpha k_\beta/k^2$ and $\tilde D_{\alpha\beta} = \sum_{\bm k} \tilde{\cal  D}_{2,\bm k} k_\alpha k_\beta/k^2$ have common principal axes $x$, $y$, $z$.}

Similarly to Eq.~(\ref{3terms}), for complete description of the fine structure, we also include a diagonal matrix describing the singlet--triplet splitting, ${\cal H}_{ST} = {\rm diag}(E_{ST},0,0,0)$ [cf. Eq.~\eqref{EST:mat}], and the matrix of the isotropic electron-hole short-range exchange interaction [cf. Eqs.~\eqref{iso:mat:hh} and \eqref{delta:iso}]:
\begin{equation}
\label{ISO:mat:Simple}
{\cal H}_{\rm short}=\begin{pmatrix}
3\delta_{\rm iso} & {\tilde\delta_{\rm iso}} & 0& {\sqrt{2}\tilde\delta_{\rm iso}} \\
{\tilde\delta_{\rm iso}} & 3\delta_{\rm iso} & 0 & -\sqrt{2}\delta_{\rm iso}\\
0 & 0 & 4\delta_{\rm iso} & 0\\
{\sqrt{2}\tilde\delta_{\rm iso}} & -\sqrt{2}\delta_{\rm iso} & 0& 2\delta_{\rm iso}
\end{pmatrix}.
\end{equation}
%\sMisha{There is no mixing between the sets $|1\rangle, \ldots |4\rangle$ and $|\bar 1\rangle, \ldots, |\bar 4\rangle$, in accordance with time-reversal symmetry.}

\begin{figure}
\includegraphics[width=\linewidth]{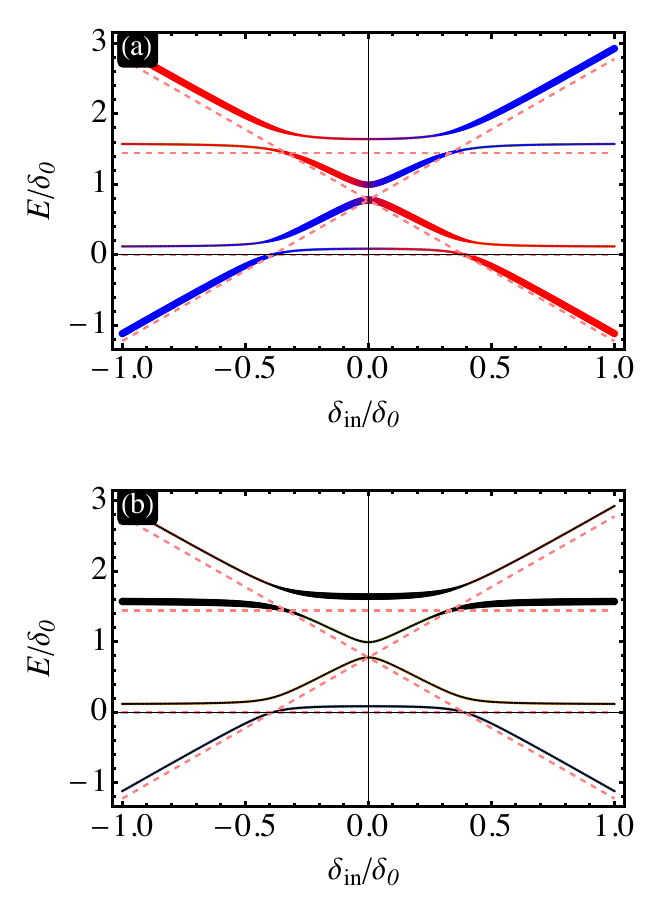}
\caption{Energy spectrum of excited trion states within simple $\Gamma_6 \times \Gamma_7$ bands calculated in the \textbf{model I.2}, Eq.~\eqref{3:param}, with $E_{ST}=\delta_0/2$, $\delta_{\rm out}=-\delta_0/6$.  (a) Emission intensity (line thickness) and linear polarization in $x$--$y$ axes frame (color from blue corresponding to the $x$-polarization to red corresponding to the $y$-polarization) superimposed on the energy spectrum. (b) Emission intensity in the $z$ polarization (line thickness) superimposed on the energy spectrum. Thin red dashed lines show asymptotics at $E_{ST}=0$. The energy is reckoned from the triplet trion state in the absence of exchange interaction.}\label{fig:model2:simple}
\end{figure}

For the further analysis we resort to the \textbf{model I.2} (see Sec.~\ref{subsec:hh:tr}) where we neglect the overlap integrals between the $1s$ and $2p_h$ states. Thus, $\tilde \delta_{\rm iso} = -\delta_{\rm iso}$ and $\tilde{D}_{2,\bm k} = -{D}_{2,\bm k}$ [cf. Eq.~\eqref{model2}], and the interaction can be conveniently parametrized by three values
\begin{subequations}
\label{3:param}
\begin{align}
&\delta_0 = 4\delta_{\rm iso} + \frac{4\pi}{3\varepsilon_b} \sum_{\bm k} {D}_{2,\bm k},\\
&\delta_{\rm in} = \frac{\pi}{\varepsilon_b} \sum_{\bm k}{D}_{2,\bm k} \frac{k_x^2 - k_y^2}{k^2},\\
&\delta_{\rm out} = \frac{\pi}{\varepsilon_b} \sum_{\bm k} {D}_{2,\bm k} \frac{ \frac32 k_\parallel^2 - k^2}{k^2}.
\end{align}
\end{subequations}
The parameter $\delta_0$ accounts for the isotropic exchange interaction, the parameters $\delta_{\rm in}$ and $\delta_{\rm out}$ characterize the in-plane and out-of-plane anisotropy respectively. The last parameter can be also recast as $\delta_{\rm out} =  (3\pi/2\varepsilon_b) \sum_{\bm k} {D}_{2,\bm k} {(k^2/3 - k^2_z)}/{k^2}$.

Numerically calculated energy spectrum, emission intensity and linear polarization are depicted in Fig.~\ref{fig:model2:simple}. Generally, all four states are optically active. The in-plane anisotropy $\delta_{\rm in}$ introduces a linear polarization in the $(x$,$y)$ plane and the splitting of the states. At large values of $\delta_{\rm in}$, there are two distinct states active in the $x$ and $y$ in-plane polarizations, Fig.~\ref{fig:model2:simple}(a), and a third state active in $z$-polarization whereas the fourth state becomes dark. 

As in the heavy-hole case, we can also trace the transition to the exciton limit by formally removing the resident ($1s$) electron from the system. To that end we set $E_{ST}=0$, $\tilde{D}_{2,\bm k}  = -{D}_{2,\bm k}$, and $\tilde \delta_{\rm iso} = -\delta_{\rm iso}$. Then one spin-degenerate dark state of the ``extra'' electron remains, with energy $E_{\rm dark} =  0$, and there is a triplet of two-fold degenerate (in the remote electron spin) states polarized along the principal axes of the quantum dot with energies
\begin{subequations}
\label{exciton:triplet:-}
\begin{align}
&E_x = \delta_0 + \frac{4}{3}\delta_{\rm out} + 2 \delta_{\rm in}, \\
&E_y =\delta_0 + \frac{4}{3}\delta_{\rm out} -2 \delta_{\rm in}, \\
&E_z = \delta_0 - \frac{8}{3}\delta_{\rm out}. \hspace{1.2 cm}
\end{align}
\end{subequations}
Naturally,  $4{\delta_{\rm in}}$ describes the splitting of the $x$- and $y$-polarized bright exciton states, $-4\delta_{\rm out}$ describes the splitting between the $z$-polarized state and the midpoint energy of $x$--$y$-polarized doublet, and $\delta_0$ is the bright-dark splitting that includes both short-range and isotropic long-range parts. The energies of the optically active triplet can be recast in somewhat different form highlighting the effect of the exchange interaction on the energy and polarization of the state
\begin{subequations}
\label{exciton:triplet}
\begin{eqnarray}
&&E_x = 4\delta_{\rm iso} + \frac{4\pi}{\varepsilon_b} \sum_{\bm k} |{D}_{\bm k}^{2p,2p_h}|^2 \frac{k_x^2}{k^2}, \\
&& E_y = 4\delta_{\rm iso} + \frac{4\pi}{\varepsilon_b} \sum_{\bm k} |{D}_{\bm k}^{2p,2p_h}|^2 \frac{k_y^2}{k^2}, \\
&&E_z = 4\delta_{\rm iso} + \frac{4\pi}{\varepsilon_b} \sum_{\bm k} | {\cal }D_{\bm k}^{2p,2p_h}|^2 \frac{k_z^2}{k^2},
\end{eqnarray}
\end{subequations}
in agreement with eigenenergies of the Hamiltonian~\eqref{exc:fine:1:simple}. In an isotropic (or cubic) quantum dot, the triplet is degenerate with the energy $$4\delta_{\rm iso} + 4\pi/(3\varepsilon_b) \sum_{\bm k} | {D}_{\bm k}^{2p,2p_h}|^2.$$

\begin{table*}[ht]
\caption{Representative exchange-induced exciton fine-structure energy ranges in quantum dots and nanocrystals.}
\label{tab:exchange-values}
\begin{ruledtabular}
%\begin{tabular}{p{0.25\textwidth}p{0.29\textwidth}p{0.30\textwidth}}
\begin{tabular}{ccp{0.30\textwidth}}
System & Bright--dark splitting $\Delta_{\rm BD}$ & Anisotropic bright-state splitting \\
\hline
In(Ga)As/GaAs QDs$^{a}$ & $\sim0.1$--$0.5$~meV &  $\sim10$--$100~\mu$eV, up to few hundred~$\mu$eV \\
GaAs/AlGaAs excited (hot) complexes$^{b}$ & $\Delta_{\rm BD} \sim0.1$--$0.2$~meV; $E_{ST}\sim5$~meV & same order as the isotropic e--h exchange; can be comparable or larger \\
CdSe/CdS core--shell NCs$^{c}$ & $\sim 0.3$--$1.9$~meV & not resolved in nearly spherical NCs$^{c}$ \\
Lead-halide perovskite NCs$^{d}$ & $\sim 1$~meV &  $\sim0.3$--$1$~meV  \\
Localized WSe$_2$ excitons$^{e}$ & $\sim 10$~meV (intravalley exciton)$^{f}$ & $\sim0.5$--$0.8$~meV \\
\end{tabular}
\end{ruledtabular}
\vspace{2pt}
%\begin{minipage}{0.96\textwidth}
\footnotesize
$^{a}$ Refs.~\cite{stevensonMagneticfieldinducedReductionExciton2006,seguinSizeDependentFineStructure2005,akopianEntangledPhotonPairs2006}; 
$^{b}$ Ref.~\cite{ivchenko302}; 
$^{c}$ Ref.~\cite{brovelliNanoengineeredElectronhole2011}; 
$^{d}$ Refs.~\cite{fuNeutralChargedExciton2017,beckerBrightTripletExcitons2018,sercelExcitonFineStructure2019,houRevealingExcitonFine2021}. Since the precise ordering of the bright and dark states in perovskite nanocrystals is under debate we quote only the characteristic energy scales; 
$^{e}$ Refs.~\cite{srivastava,wangHighlyPolarizedSingle2021}; $^{f}$ Ref.~\cite{renControlBrightDarkExciton2023}.
%\end{minipage}
\end{table*}

\section{Discussion and summary}\label{sec:disc}

We have calculated above the fine structure of the excited trion states caused by the LRE interaction in semiconductor quantum dots for two models of the band structure: simple $\Gamma_6 \times \Gamma_7$ bands and heavy-hole $\pm3/2$ valence band. We focus on isolated trion orbital state, e.g., $1s2p_x$-$2p_{h,x}$ where the photoexcited electron-hole pair occupies specific, $2p_x$-shell orbital. A possibility to consider such a state independently of other $2p$ states, e.g., the $1s2p_y$-$2p_{h,y}$ configuration, is guaranteed by the in-plane anisotropy of the quantum dot confinement potential. The latter configuration $1s2p_y$-$2p_{h,y}$ is considered similarly, the result differs in the sign of linear polarization~$P_l$.

As mentioned in Sec.~\ref{subsec:states}, the very same analysis holds for the $1s2s$-$2s_h$ excited trion with the replacement of Eqs.~\eqref{Dddd} by 
\begin{subequations}
\label{Dddd:2s}
\begin{align}
{D}_{2,\bm k} = |{D}_{\bm k}^{1s,2s_h}|^2 +|{D}_{\bm k}^{2s,2s_h}|^2, \\ 
\tilde{D}_{2,\bm k} = |{D}_{\bm k}^{1s,2s_h}|^2 - |{D}_{\bm k}^{2s,2s_h}|^2,
\end{align}
\end{subequations}
and analogous revisions of Eqs.~\eqref{delta:iso},
\begin{subequations}
\label{delta:iso:2s}
\begin{align}
\delta_{\rm iso}= C \int d\bm r |\varphi_{2s_h}(\bm r)|^2 (|\varphi_{1s}(\bm r)|^2 + |\varphi_{2s}(\bm r)|^2),\\
\tilde \delta_{\rm iso}  = C \int d\bm r |\varphi_{2s_h}(\bm r)|^2 (|\varphi_{1s}(\bm r)|^2 - |\varphi_{2s}(\bm r)|^2),
\end{align}
\end{subequations}
where ${D}_{\bm k}^{i,j}$ are introduced in Eq.~\eqref{D:1s1s}.

It is instructive to briefly address the fine structure in specific cases of nearly uniaxial or nearly isotropic quantum dots were the excited $2p_\alpha$-shells ($\alpha=x,y$, or $z$) become almost degenerate. In these cases of close-to-axially-symmetric quantum dot, as demonstrated in Ref.~\cite{ivchenko232} for excitons, the direct Coulomb interaction mixes almost degenerate states  and splits them to four terms
\begin{eqnarray}
\label{active}
&&\Psi_{xx,yy}^{(\pm)} = \frac{1s 2p_x 2p_{h,x} \pm 1s2p_y 2p_{h,y}}{\sqrt{2}},\\ && \Psi_{xy,yx}^{(\pm)} = \frac{1s 2p_x 2p_{h,y} \pm 1s2p_y 2p_{h,x}}{\sqrt{2}}. \nonumber
\end{eqnarray}
The optically active combination $\Psi_{xx,yy}^{(+)}$ is split by the Coulomb energy $V_{\rm C} \sim e^2/(\varepsilon_0 a)$ ($a$ is the quantum dot size) from the combinations $\Psi_{xx,yy}^{(-)}$ and $\Psi_{xy,yx}^{(\pm)}$
belonging to the optically dark $|m_z|=2$ orbital  manifolds. Hence, provided that $E_{ST}$ in Eq.~\eqref{EST} and the fine structure splittings are smaller than the direct Coulomb energy $V_{\rm C}$, the orbital state $\Psi_{xx,yy}^{(+)}$ can be considered as isolated.

Our theory can also be applied to the `hot' trions $X^{-*}$ or $X^{+*}$ with the configurations $1s2p_\alpha$-$1s_h$ or $1s$-$1s_h2p_{h,\alpha}$, respectively, which are observed in hot photoluminescence of charged quantum dots~\cite{cortezOpticallyDrivenSpin2002,
jovanovDirectObservationMetastable2011,bennyExcitationSpectroscopySingle2012,molasIntershellExchangeInteraction2013,molasMagneticFieldEffect2014,ivchenko302}. As a rule, the emission of $S=1$ two-electron (in the $X^{-*}$) manifold is observed as a doublet at zero magnetic field: states $|2\rangle$ ($|\bar 2\rangle$) with $|J_z| = 3/2$ and  $|3\rangle$ ($|\bar 3\rangle$) with $|J_z| = 1/2$ in Fig.~\ref{fig:hh}, while singlet states $|1\rangle$, $|\bar 1\rangle$ are sufficiently far in energy~\cite{ivchenko302}. The effective $2\times 2$ Hamiltonian is given by Eq.~\eqref{H23} where the parameters $\delta_0$ and $\delta_x$ are expressed in the form of Eqs.~\eqref{delta0x} with
\begin{subequations}
\label{Dddd:hot}
\begin{align}
{D}_{2,\bm k} = |{D}_{\bm k}^{1s,1s_h}|^2 +|{D}_{\bm k}^{2p,1s_h}|^2,\hspace{0.8 cm}\\
\delta_{\rm iso}= C \int d\bm r |\varphi_{1s_h}(\bm r)|^2 (|\varphi_{1s}(\bm r)|^2 + |\varphi_{2p}(\bm r)|^2).
\end{align}
\end{subequations}
In the absence of magnetic field the upper and lower lines of the doublet are linearly polarized (with opposite signs of $P_l$), see Fig.~\ref{fig:model1}. In the presence of magnetic field applied along the $z$-axis, each line in the doublet splits into two. In  GaAs/AlGaAs [111] possessing the $C_{3v}$ trigonal symmetry the linear polarization is absent, while the magnetic field additionally mixes bright states with dark states $|J_z|=5/2$ ($|4\rangle$, $|\bar 4\rangle$) as observed in experiment~\cite{ivchenko302}.

It is worth stressing that, in contrast to the short-range exchange interaction which is essentially isotropic and insensitive to the particular shape of the envelope functions, the LRE interaction is generally anisotropic and results in linear polarization sensitive fine structure of neutral and charged excitons.

To demonstrate that the predicted fine structure of the trion states is experimentally observable we note that the key parameters of the model: isotropic splitting $\delta_0$ and anisotropic splitting $\delta_x$ (heavy-hole model) or $\delta_{\rm in, out}$ (simple bands model) basically correspond to the exchange energies in excitons. Table~\ref{tab:exchange-values} summarizes exchange-induced exciton fine-structure splittings measured in several classes of semiconductor quantum dots and nanocrystals. In the heavy-hole model considered in Sec.~\ref{subsec:hh:tr},  $\delta_0=\Delta_{\rm BD}/2$ where $\Delta_{\rm BD}$ is the bright--dark splitting and $|\delta_x|=\Delta_{xy}/2$ for the anisotropic splitting $\Delta_{xy}$ of the radiative doublet, see Eq.~\eqref{limit:doublets}. Thus, for conventional III--V quantum dots, the literature values correspond approximately to $\delta_0\sim 0.05$--$0.25$~meV and, typically, $|\delta_x|\sim 0.005$--$0.05$~meV, although substantially larger anisotropic splittings are found in strongly anisotropic dots. The electron--hole exchange constants involving excited $p$-shell states of about $0.09$--$0.19$~meV have been extracted in GaAs/AlGaAs quantum dots~\cite{ivchenko302}. The electron--electron singlet--triplet splitting $E_{ST}$ is generally larger; a characteristic value $E_{ST}\sim5$~meV was reported for hot trions in the same class of structures~\cite{ivchenko302}. In the simple-band model, $\delta_0=\Delta_{\rm BD}$, while $4\delta_{\rm in}=E_x-E_y$ and $-4\delta_{\rm out}=E_z-(E_x+E_y)/2$, so that the sub-meV to meV anisotropic splittings observed in lead-halide perovskite nanocrystals correspond to anisotropy parameters on the $0.1$~meV scale. For localized excitons in WSe$_2$, zero-field fine-structure splittings of order $0.5$--$0.8$~meV have been reported. Importantly, polarization-resolved single-dot spectroscopy can resolve splittings in the tens-of-$\mu$eV range: for example, a splitting of $27\pm3~\mu$eV was resolved in InGaAs/GaAs dots~\cite{akopianEntangledPhotonPairs2006}, while characteristic separations of $100$--$150~\mu$eV were observed for excited charged complexes in GaAs/AlGaAs dots~\cite{ivchenko302}. Hence, the exchange-induced spectral structures discussed here lie within the experimentally accessible range.

Naturally, the long-range exchange interaction is sensitive to the shape and anisotropy of quantum dots. Hence, it can be modified by variation of the quantum dot shape and size. Generally, smaller quantum dots have higher electron-hole and electron-electron overlaps and, consequently, larger exchange interaction, and variation of quantum dot anisotropy degree can be used to tune anisotropic splittings. Here, the sensitivity to the dot shape results from the anisotropic contributions in Eqs.~\eqref{delta0x} and \eqref{3:param}. Furthermore, similarly to the case of neutral exciton fine structure, a tuning of the anisotropic part of the LRE interaction can be performed by application of magnetic~\cite{stevensonMagneticfieldinducedReductionExciton2006,ivchenko233} and electric fields~\cite{kowalikInfluenceInplaneElectric2005,gerardotManipulatingExcitonFine2007} or elastic strain~\cite{tartakovskiiEffectThermalAnnealing2004,abbarchiExcitonFineStructure2008} with the latter effect being expected to be most pronounced transition-metal dichalcogenide systems~\cite{glazovExcitonFineStructure2022}.

Now we discuss the relation between the electron-hole LRE interaction-induced fine structure of the excited trion states in quantum dots studied here and the  anisotropic splitting of the Fermi-polarons (also known as Suris tetrons) in two-dimensional transition metal dichalcogenides uncovered in the recent works~\cite{iakovlevFermiPolaronFine2023,iakovlevLongitudinaltransverseSplittingFine2024, yagodkinFermiPolaronsStraininduced2025}. While both effects stem from the electron-hole exchange interaction, the manifestations are quite different. In the trions, the LRE interaction only partially lifts the degeneracy of the three-particle trion leaving each state two-fold Kramers degenerate since we deal with the half-integer spin complex. In the Fermi-polarons/tetrons, the exciton complex effectively comprises four fermions (trion and Fermi-sea hole) and possesses the integer spin. As a result, anisotropy of the system in combination with the exchange interaction fully lifts the degeneracy of the optically active states.

To conclude, we have developed a theory of fine structure of excited negative trions in semiconductor quantum dots controlled by the long-range part of the electron-hole exchange interaction. In contrast to the ground singlet trion, whose two-fold spin degeneracy is protected by the time-reversal symmetry and which  acquires only an exchange-induced Lamb shift, the excited trions demonstrate non-trivial fine structure related to the competition of electron-electron and electron-hole exchange interactions that mixes the singlet and triplet configurations. 

The fine structure is calculated for both heavy-hole trions where the optical transitions are possible only in the in-plane polarization and for the simple band structure where the trions are also active in the $z$-polarization. We have traced a transition to the limit where the resident electron is effectively removed from the quantum dot. In this case, the fine structure of excited trions reduces to that of optically active excitons. As opposed to the isotropic short-range exchange interaction, the long-range exchange interaction is highly sensitive to the anisotropy of the form of envelope functions of electron-hole complexes and, hence, the quantum dot shape. It is the long-range exchange interaction that produces linear polarization of exchange-split excited trion states.

\acknowledgements

The authors are grateful to K. Bolotin, T.V. Shubina, A.A. Toropov, G. Veyshtort, and Z.A. Iakovlev for valuable discussions. This work was supported by the RSF project 23-12-00142-Continuation.

\appendix
\section{Microscopic derivation of the long-range exchange interaction}\label{app:exch}

Here we derive Eq.~\eqref{H:exch} directly from the Coulomb interaction
between the conduction- and valence-band electrons. We consider the
initial and final electron--hole pair states
\begin{subequations}
\begin{align}
|A'\rangle &=
a^\dagger_{s'_e,\bm k/2+\bm q'}
b^\dagger_{s'_h,\bm k/2-\bm q'}|0\rangle ,
\label{A:initial}
\\
|A\rangle &=
a^\dagger_{s_e,\bm k/2+\bm q}
b^\dagger_{s_h,\bm k/2-\bm q}|0\rangle ,
\label{A:final}
\end{align}
\end{subequations}
where $|0\rangle$ corresponds to the completely filled valence band. It is convenient to work in the electron representation where the hole operators $b^\dag_{s,\bm k}$ ($b_{s,\bm k}$) and the valence band electron operators $c^\dag_{s,\bm k}$ ($c_{s,\bm k}$) are related by~\cite{birSymmetryStraininducedEffects1974}
\begin{equation}
\label{time:reversal}
b^\dag_{s,\bm k} = c_{-s,-\bm k}, \quad b_{s,\bm k} = c^\dag_{-s,-\bm k},
\end{equation}
in agreement with the general time-reversal symmetry requirement at the electron-hole to electron-electron representation change.
The single-electron Bloch functions corresponding to these operators are written as
\begin{equation}
\psi_{n s,\bm k}(\bm r)
=
\frac{1}{\sqrt{\mathcal V}}
e^{\mathrm i\bm k\bm r}
u_{n s,\bm k}(\bm r),
\label{bloch:general}
\end{equation}
where $n=c,v$ denotes the conduction or valence band, respectively, $s$ is the spin quantum number,
$u_{n s,\bm p}(\bm r)$ is the periodic Bloch amplitude, and $\mathcal V$ is the
normalization volume.

The Coulomb interaction between electrons can be written as
\begin{equation}
\hat H_C
=
\frac{1}{2}
\int d\bm r_1 d\bm r_2\,
\hat\Psi^\dagger(\bm r_1)
\hat\Psi^\dagger(\bm r_2)
\frac{e^2}{\varepsilon_b|\bm r_1-\bm r_2|}
\hat\Psi(\bm r_2)
\hat\Psi(\bm r_1),
\label{HC:real}
\end{equation}
where $\hat{\Psi}(\bm r)$ and $\hat{\Psi}^\dag(\bm r)$   are the second quantized annihilation and creation operators of electrons in the real space representation.
The matrix element
$\langle A|\hat H_C|A'\rangle$ contains the direct electron--hole
Coulomb interaction and the exchange contribution. The latter has the
form~\cite{pikusExchangeInteractionExcitons1971,birSymmetryStraininducedEffects1974}
\begin{multline}
V^{\rm exch}_{AA'}
=
\int d\bm r_1 d\bm r_2\,
\psi^*_{c s_e,\bm p_e}(\bm r_1)
\psi_{v s_v,\bm p_v}(\bm r_1)
\\
\times
\frac{e^2}{\varepsilon_b|\bm r_1-\bm r_2|}
\psi^*_{v s'_v,\bm p'_v}(\bm r_2)
\psi_{c s'_e,\bm p'_e}(\bm r_2),
\label{Vex:real}
\end{multline}
where [cf. Eq.~\eqref{time:reversal}]
\begin{subequations}
\label{wavevectors}
\begin{align}
&\bm p_e = \bm k/2 + \bm q, \quad &\bm p_e' = \bm k/2 + \bm q',\\
&\bm p_v = -\bm k/2 + \bm q, \quad & \bm p_v' = -\bm k/2 + \bm q',
\end{align}
\end{subequations}
and we use the background dielectric constant relevant for the screening of the long-range exchange contribution~\cite{kiselevDielectricTheorySemiconductors1974}. Using standard Fourier decomposition
\begin{equation}
\frac{e^2}{\varepsilon_b|\bm r|}
=
\frac{1}{\mathcal V}
\sum_{\bm Q}
\frac{4\pi e^2}{\varepsilon_b Q^2}
e^{\mathrm i\bm Q\bm r},
\label{coulomb:fourier}
\end{equation}
we can write Eq.~\eqref{Vex:real} as
\begin{equation}
V^{\rm exch}_{AA'}
=
\frac{4\pi e^2}{\varepsilon_b\mathcal V}
\sum_{\bm Q}
\frac{
I_{cv}(\bm Q)
I_{vc}(-\bm Q)
}{Q^2},
\label{Vex:Q}
\end{equation}
where
\begin{multline}
I_{cv}(\bm Q)
=
\int d\bm r\,
\psi^*_{c s_e,\bm p_e}(\bm r)
e^{i\bm Q\bm r}
\psi_{v s_v,\bm p_v}(\bm r)\\
=
\frac{1}{\mathcal V}
\int d\bm r\,
u^*_{c s_e,\bm p_e}(\bm r)
u_{v s_v,\bm p_v}(\bm r)
e^{\mathrm i(\bm Q-\bm p_e+\bm p_v)\bm r}.
\end{multline}
Since the product of the periodic Bloch amplitudes is lattice periodic,
the integral is nonzero provided
\begin{equation}
\bm G = \bm Q-\bm p_e+\bm p_v= \bm k + \bm Q,
\end{equation}
where $\bm G$ is a reciprocal-lattice vector. Consequently, the exchange interaction can be represented as a sum over
reciprocal-lattice vectors. The contribution with $\bm G=0$ is the
long-range exchange interaction, whereas terms with $\bm G\ne0$
describe the short-range part.

The long-range exchange contribution contains the
interband overlap integral
\begin{equation}
M_{cv}
=
\left\langle
u_{c s_e,\bm q+\bm k/2}
\middle|
u_{v s_v,\bm q-\bm k/2}
\right\rangle.
\label{Mcv:def}
\end{equation}
For small $k,q \ll G$ it can be expanded to the first order in $\bm k$ with the result
\begin{multline}
M_{cv}
=
\frac{k_\alpha}{2}
\left[
\left\langle
\frac{\partial u_c}{\partial q_\alpha}
\middle|u_v
\right\rangle
-
\left\langle
u_c\middle|
\frac{\partial u_v}{\partial q_\alpha}
\right\rangle
\right]
\\
=
-k_\alpha
\left\langle
u_c\middle|
\frac{\partial u_v}{\partial q_\alpha}
\right\rangle .
\label{Mcv:expand}
\end{multline}
Here and below summation over repeated Cartesian indices ($\alpha=x,y,z$) is implied and the derivatives are evaluated at $\bm k, \bm q=0$. In
the second line we used the relation which can be readily verified: ${\partial}
\langle u_c|u_v\rangle /{\partial q_\alpha} =0$. The derivative of the Bloch amplitude is related to the interband
dipole matrix element. Introducing
\begin{equation}
d_{\alpha;s_e,s_h}
=
\mathrm ie
\left\langle
u_{c,s_e}
\middle|
\frac{\partial u_{v,s_v}}{\partial q_\alpha}
\right\rangle ,
\qquad
s_v=-s_h,
\label{dipole:berry}
\end{equation}
with the appropriate phase factors implied by the time-reversed
definition of the hole states, Eq.~\eqref{Mcv:expand} becomes
\begin{equation}
M_{cv}
=
\frac{i}{e}
k_\alpha d_{\alpha;s_e,s_h},
\label{Mcv:dipole}
\end{equation}
and
\begin{equation}
e^2 M_{cv}M_{vc}
=
k_\alpha k_\beta
d_{\alpha;s_e,s_h}
d^*_{\beta;s'_e,s'_h}.
\label{product:dipoles}
\end{equation}
Substitution into Eq.~\eqref{Vex:Q} gives the long-range part of the
exchange matrix element,
\begin{equation}
V^{\rm LR}_{AA'}
=
\frac{4\pi}{\varepsilon_b\mathcal V}
\frac{k_\alpha k_\beta}{k^2}
d_{\alpha;s_e,s_h}
d^*_{\beta;s'_e,s'_h}.
\label{VLR:AA}
\end{equation}
Returning to the electron--hole representation, the corresponding
second-quantized operator is (we set $\mathcal V \equiv 1$)
\begin{multline}
\hat H_{\rm exch}
=
\frac{4\pi}{\varepsilon_b}
\sum_{\bm k,\bm q,\bm q'}
\frac{k_\alpha k_\beta}{k^2}
d_{\alpha;s_e,s_h}
d^*_{\beta;s'_e,s'_h}
\\
\times
a^\dagger_{s_e,\bm k/2+\bm q}
b^\dagger_{s_h,\bm k/2-\bm q}
b_{s'_h,\bm k/2-\bm q'}
a_{s'_e,\bm k/2+\bm q'} .
\label{H:exch:microscopic}
\end{multline}
in agreement with  Eq.~\eqref{H:exch} of the main text. Note that Eq.~\eqref{H:exch:microscopic} is also consistent with the form of the exchange interaction derived previously for two-dimensional excitons in transition metal dichalcogenides, see Refs.~\cite{glazovExcitonFineStructure2014,yuDiracConesDirac2014}. In this case, one needs to sum Eq.~\eqref{H:exch:microscopic} over $k_z$ to account for two-dimensional nature of the system. 

The equivalence of two approaches reflects a remarkable property of physics, where, at times, two different rigorous considerations lead to identical results.

\end{document}